\documentclass[aps,twocolumn,amsmath,amssymb,preprintnumbers]{revtex4}
\usepackage{amsmath} \usepackage{amsfonts} \usepackage{amssymb}
\usepackage{bbm}
\usepackage{epsfig}
\usepackage{graphics}
\usepackage{graphicx}
\newcommand{\be}{\begin{equation}}
\newcommand{\ee}{\end{equation}}
\newcommand{\ba}{\begin{eqnarray}}
\newcommand{\ea}{\end{eqnarray}}
\newcommand{\nn}{\nonumber}
\newcommand{\kr}{\rangle}
\newcommand{\kl}{\langle}

\newcommand{\cD}{{\cal D}}

\newcommand{\cP}{{\cal P}}

\newcommand{\tcu}{\tilde{\cal U}}
\newcommand{\tu}{\tilde U}

\newcommand{\ns}{{\cal N}_s}

\begin{document}

\title[ ]{Probabilistic time}

\author{C. Wetterich}
\affiliation{Institut  f\"ur Theoretische Physik\\
Universit\"at Heidelberg\\
Philosophenweg 16, D-69120 Heidelberg}

\begin{abstract}
The concept of time emerges as an ordering structure in a classical statistical ensemble. Probability distributions $p_\tau(t)$ at a given time $t$ obtain by integrating out the past and future. We discuss all-time probability distributions that realize a unitary time evolution as described by rotations of the real wave function $q_\tau(t)=\pm \sqrt{p_\tau(t)}$. We establish a map to quantum physics and the Schr\"odinger equation. Suitable classical observables are mapped to quantum operators. The non-commutativity of the operator product is traced back to the incomplete statistics of the local-time subsystem. Our investigation of classical statistics is based on two-level observables that take the values one or zero. Then the wave functions can be mapped to elements of a Grassmann algebra. Quantum field theories for fermions arise naturally from our formulation of probabilistic time. 

\end{abstract}

\maketitle

\section{Introduction}
\label{Introduction}
What is time? This question touches a basic concept in physics. In many theoretical formulations, such as mechanics or quantum field theory, time appears as an ``a priori notion''. Time looses its absolute character by the unification of space and time in special and general relativity. Still, spacetime is assumed to exist a priori in these theories, although time intervals and distances depend now on the particular inertial system and metric and therefore become dynamical rather than a priori properties. We pursue in this paper the notion that time (or more generally spacetime) arises as an effective concept in a theoretical setting which does not postulate its existence from the beginning. More precisely, we explore the possibility that time emerges  as an ordering structure in a probabilistic setting based on a classical statistical ensemble. As such, it is not distinguished from other possible structures among the observables and appears as a derived quantity. Not all possible probability distributions for the states of the ensemble are compatible with the concept of time. Only a subclass of the probability distributions allows the existence of time, whereas for others ``there is no time''. Such a setting opens the logical possibility that time could appear as an approximate concept. It provides a formal framework for discussions of the idea that the notion of time may loose its validity under extreme circumstances as the ``beginning of the universe.''

Many thoughts have been given in the past to the idea of an ``emergence of time'' from a more fundamental setting \cite{A,B,2A,AA,AB,4A,AC,R6,6A,R1,R2,R3,R4,10A,R5,11A}. Our approach is close to general statistics \cite{A}, where the emergence of space and geometry has been discussed previously \cite{B}. Since the basic concept of a statistical ensemble contains no notion of a different signature of the metric for space and time, this crucial difference between space and time has to be traced back to a difference in the relevant structures among observables. We will see that the particularity of time is associated to the presence of a  ``unitary evolution law'' which singles out time as compared to space.

The starting point of our approach is the description of Nature in terms of a classical statistical ensemble with positive probabilities. Observables have fixed values in the ``classical states'' of the ensemble. The association of the abstract observables with real observations becomes only possible once appropriate general structures among observables are identified - those structures include time and space. We will find that quantum physics and quantum field theory emerge naturally in our approach. Once these structures are identified we can proceed to an association between quantum operators and observations. In this paper we remain on a rather abstract level. Many different associations of physical observations which observables in a classical ensemble describing, for example, two-state quantum mechanics can be made. A special role is played by the Hamiltonian operator which is directly related to the concept of time-evolution.

While quantum physics is always related to a probabilistic description of Nature, we find in our approach that all quantum features can be described {\em only} in terms of positive ``classical'' probabilities. In the usual view the quantum wave function $\psi$ carries more information than the probabilities which are associated to $|\psi|^2$. This additional information is stored in the phase of the wave function. It is therefore widely believed that quantum physics cannot be expressed in terms of positive probabilities alone. In contrast, our investigation reveals that all quantum features can indeed be described in terms of classical probabilities. In our context this is closely related to the emergence of time. The quantum features of a wave function and the non-commuting operators at a given time arise from a classical statistical ensemble for all times. At a given time the statistical description is incomplete since the statistical information concerning the past and future is ``integrated out''. 

We work within a general probabilistic setting where one assumes the existence of one reality, while physical laws allow us only to compute probabilities. (Deterministic laws are special cases for probabilities extremely close to one or zero.) Such a setting may be considered as fundamental (probabilistic realism). However, our discussion also covers macroscopic statistical systems where an effective concept of evolution and time is not necessarily directly related to fundamental time. 

As a simple example we may consider an Ising-type model on a two-dimensional square lattice with discrete lattice points labeled by $x^0$ and $x^1$. For each lattice site one has an Ising spin that can only take two values, or an associated occupation number $n(x^0,x^1)$ that takes the values zero or one. The possible association with bits is obvious. The states $\omega$ are configurations of Ising spins or ordered sequences of bits. We consider a classical ensemble with one given probability distribution $\{p_\omega\}$. More formally, we associate to each state $\omega$ a positive number $0\leq p_\omega\leq 1$, with normalization $\sum_\omega p_\omega =1$. 

We ask the question for which type of $\{p_\omega\}$ one of the lattice directions, say the direction of $x^0$, will appear as a time direction, and the other, say the direction of $x^1$, as a space direction. The structures of time and space can be linked to the behavior of correlation functions \cite{B}. We may consider connected correlation functions in different directions, as
\ba\label{A1}
G_0(d;x)&=&\kl n(x^0+d,x^1)n(x^0,x^1)\kr_c,\nn\\
G_1(d;x)&=&\kl n(x^0,x^1+d)n(x^0,x^1)\kr_c.
\ea
For example, time structures could be periodic, resulting in a periodic correlation function $G_0(nd_0)=G_0(0)$, for a suitable $d_0$ and integer $n$, while space structures may be reflected by an exponential or power law decay of $G_1(d)$ for large $|d|$. We want to formalize these different types of structures and find out what type of probability distributions $\{p_\omega\}$ realize a time structure. Time will be associated with appropriate properties of an ``evolution law'' that allows the computation of effective probabilities $\{p_\tau(x^0)\}$ for some ``time'' $t=x^0$, and relates $\{p_\tau(x^0+\epsilon)\}$ to $\{p_\tau(x^0)\}$. 

The statistical ensemble described by the probability distribution $\{p_\omega\}$ may be called the ``all-time ensemble''. It permits to compute expectation values and correlations for occupation numbers at arbitrary times. Knowledge of $\{p_\omega\}$ would contain the information about measurements or events at all times - present, past and future. In contrast, a local (in time) probability distribution $\{p_\tau(t)\}$ contains the information about correlation functions at a given time $t$. A sequence of local probability distributions $\{p_\tau(t)\}$ for all $t$ may be associated with the state of a ``local-time subsystem''. The sequence of $\{p_\tau(t)\}$ contains much less information than $\{p_\omega\}$. The transition from the all-time ensemble to a state of the local-time subsystem therefore involves an important ``coarse graining of the information''. This coarse graining leads to incomplete statistics \cite{3}, \cite{QPSS} and is responsible for many of the characteristic conceptual features of quantum mechanics. 

The ``local-time probability distribution'' $\{p_\tau(t)\}$ and its time evolution shows analogies to the transfer matrix formulation for the Ising model. However, in the conventional two-dimensional Ising model both $x^0$ and $x^1$ are equal footing. The correlation functions typically decay for large $|d|$ such that we may associate both $x^0$ and $x^1$ with space directions. In the present paper we investigate generalized Ising-type models where the all-time ensemble $\{p_\omega\}$ differs from the standard Ising model. We concentrate on all-time ensembles that lead to an evolution law which allows for periodic structures, such that $x^0$ can be associated with a time direction rather than a space direction. Perhaps surprisingly, a large class of suitable $\{p_\omega\}$ is found to describe a quantum field theory for fermions in one time and one space dimension.

The Ising model is just one particular example and we will keep our discussion of time structures more general. It is rather straightforward to implement in a statistical ensemble the structure of an {\em ordered chain}. It is sufficient that a class of observables exists such that all observables in this class depend on some label $t_i$, and that every $t_i$ has only two neighboring labels $t_{i-1}$ and $t_{i+1}$. This  allows us to order the observables by an increasing label $t_i$ of their ``time argument''. (The direction of time'' plays no role at this level. There may also exist other observables that have no ``time label'' $t_i$.) We will consider here the basic setting of two-level observables $N_\alpha(t_i)$ that can only take values one or zero, where the index $\alpha$ is a collective index for other properties as location in space or quantum numbers of particles. More complicated observables can always be reduced to a sufficient number of such yes/no decisions. For the Ising model discussed above $t_i$ can be associated to values of the lattice coordinate $x^0$, while $\alpha$ corresponds to values of the lattice coordinate $x^1$. 

We assume that the observables $N_\alpha(t_i)$ take definite values for every state of the statistical ensemble. Thus every state can be labeled by all the values $n_\alpha(t_i)=0,1$ that the observables $N_\alpha(t_i)$ take in this state. We also assume that the sequence of numbers $\big \{n_\alpha(t_i)\big \}$ is sufficient to label the states of the classical statistical ensemble, which we will denote by $\omega=\{n_\alpha(t_i)\}$. Furthermore, we assume  that the range of $\alpha$ is the same for all $t_i$. Without loss of generality we can use labels $t_i=\epsilon i, ~i\in {\mathbbm Z}$, such that the chain consists of equivalent points in the interval $t_{in}\leq t_i\leq t_f$. Taking the limit $\epsilon\to 0$ at fixed $t_{in},t_f$ results in observables depending on a continuous time-coordinate $t$. (We could have started our discussion with a continuous time label from the beginning. However, the limiting procedure of a discrete setting allows for more conceptual clarity since all quantities are well defined for arbitrarily small but nonzero $\epsilon$.)

At this stage nothing distinguishes time from similar {\em ``one-dimensional continuous ordering structures''} where observables depend on some real parameter, as a space coordinate, momentum or energy. For all one-dimensional ordering structures one can define the notion of {\em ``local probabilities''} $p_\tau(t)$. (In a more general context, the label $t$ is then replaced by a space coordinate $x$ or an energy $E$.) While the general probability distribution associates to each state $\omega=\big \{n_\alpha(t_i)\big \}$ a positive real number $p_\omega$, with$\sum_\omega p_\omega=1$, the local probabilities depend only on the sequence of ``local occupation numbers'' $\tau=[n_\alpha(t)]$ at a given ``time'' $t$. the local probability $p_\tau(t)$ obtains by summing for given values of $\tau$ over all probabilities $p_\omega$ for states that share the same value of $\tau$. In other words, we consider in the generalized parameter space spanned by $(\alpha,t)$ a hypersurface at fixed $t$, and we ``integrate out'' all information concerning the ``left and right'' or the ``past and future'' of this hypersurface. By construction, the local probabilities are normalized for every $t,\Sigma_\tau p_\tau(t)=1$. 

A central property of statistical ensembles which admit a time structure is the existence of an {\em ``evolution law''} which allows the computation of $p_\tau(t)$ if $p_\tau(t-\epsilon)$ is known. (Instead of the ``immediate past'' at $t-\epsilon$ we can also generalize to a larger range of ``past probabilities''.) The evolution law is the basis for differential ``evolution equations'' for the expectation values of observables as, for example, the Schr\"odinger or Heisenberg equations in quantum mechanics. Indeed, one can define ``local observables'' which involve combinations of $N_\alpha(t)$ at a given time $t$.  Their expectation values can be computed from $p_\tau(t)$ without additional information from the past and future. An evolution law allows predictions for future expectation values if the present ones are known. There exists a variety of statistical systems which admit an evolution law - for lattice systems this issue is related to the concept of transfer matrices. At this stage we still have not yet addressed the particular properties of ``time'' as compared to some space direction. 

The difference between time and space resides in the particular properties of the evolution law. (More precisely, the properties of the evolution law are responsible for a different signature of the metric for time and space.) This paper therefore addresses a particular form of the evolution law that we call {\em ``unitary evolution law''}. A glance to the characteristics of time evolution in nature reveals the omnipresence of oscillation phenomena, as in quantum  physics or clocks. The evolution law should therefore be compatible with a periodic time evolution of the local probabilities $p_\tau(t)$. This eliminates evolution laws which lead to a monotonic behavior of $p_\tau(t)$ as, for example, the approach to a fixed point. 

For a realization of unitary evolution laws we introduce the concept of a real {\em ``wave function''} $q_\tau(t)$ in classical statistics. It equals the square root of $p_\tau(t)$ up to a sign, $p_\tau(t)=q^2_\tau(t)$. Since $q_\tau(t)$ is a vector with unit length, the periodicity in the time evolution can be accounted for naturally by rotations of this vector. A unitary time evolution acts linearly on $q_\tau(t)$, i.e. 
\be\label{A}
q_\tau (t)=\sum_\rho R_{\tau\rho}q_\rho(t-\epsilon),
\ee
with rotation matrix $R$ independent of $q$. This is the simplest evolution law which is compatible with periodic $p_\tau(t)$, as compared to the more general  non-linear case where $R$ depends on $q$ or to evolution laws involving information beyond the immediate past $t-\epsilon$. We will see that this simple assumption leads directly to the quantum formalism and the Schr\"odinger equation. Quantum physics arises naturally from the formulation of an evolution law for ``probabilistic time''. We find here an explicit example for the emergence of quantum physics from a classical statistical ensemble \cite{QPSS}. 

One may ask for which overall probability distributions $p_\omega=p\big(\big \{n_\alpha(t_i)\big\}\big)$ the unitary evolution law \eqref{A} holds. We show that for every given sequence of local probabilities $\{p_\tau(t)\}$ a large equivalence class of different $\{p_\omega\}$ exists which yields this sequence by integrating out the past and future. The question if some particular all-time ensemble $\{p_\omega\}$ is particularly natural is left to future investigations. This issue is not crucial for the investigations of the present paper. 

We use the analogy of the sequence $\big\{ n_\alpha(t_i)\big\}$ with occupation numbers for fermions in order to construct a representation of the probability density $\{p_\tau(t)\}$ in terms of Grassmann variables. If the index $\alpha$ contains a space coordinate and ``internal quantum numbers'', the Grassmann algebra describes a quantum field theory for fermions. It is remarkable how many basic concepts of fundamental physics - quantum theory, Schr\"odinger equation, fermions, Grassmann variables and quantum field theories - arise in a natural way from the attempt to formulate a probabilistic concept of time. 

This paper is organized as follows. In sect. \ref{all-time} we discuss the all-time ensemble and introduce as basic building blocks the two-level observables $N_\alpha(t_i)$. They correspond to yes/no alternatives, carrying one bit of information, or to occupation numbers with values $0,1$. Observables of this type are defined at every point $t_i$ of an ordered chain and we characterize the states $\omega$ of the classical statistical ensemble by the values of these observables $n_\alpha(t_i)$, i.e. $\omega=\big\{n_\alpha(t_i)\big\}$. The ensemble is specified by the probability $p_\omega$ for all states $\omega$. This minimal setting will be our only basic conceptual assumption. All correlation functions for the observables $N_\alpha(t_i)$ are well defined by this setting. The concept of time emerges if the probability distribution $\{p_\omega\}$ has the characteristic properties mentioned above.

We define in sect. \ref{timesubsystem} the ``local-time probability distribution'' $p_\tau(t)$. It is local in time and obtains by ``integrating out'' the past and the future, i.e. summing over all $t'\neq t$. The sequence of $\{p_\tau(t)\}$ for different $t$ specifies a state of the local-time subsystem. We also discuss simple examples for all-time probability distributions $\{p_\omega\}$ that lead to periodic local probabilities. One of these examples will later turn out to describe two-state quantum mechanics. We show that the map from the all-time ensemble to the local-time subsystem introduces equivalence classes of probability distributions. Two all-time probability distributions are equivalent if they lead to the same state of the local-time subsystem. We argue that only the information contained in the local-time subsystem is available for a description of Nature.

In sect. \ref{Evolutionlaw} we formulate an essential property of the probability distribution $\{p_\omega\}$ which allows for the structures of time and evolution: it is the existence of an evolution law which relates $p_\tau(t)$ to $p_\tau(t-\epsilon)$ for two neighboring ``time points''. In particular, we discuss the ``unitary time evolution'' which amounts to a rotation of the vector $q_\tau(t)$, where $p_\tau=q^2_\tau$. Sect. \ref{Quantum formalism} establishes the close analogy to the time evolution in quantum mechanics. We construct an explicit map from the local-time subsystem to a quantum system. 

In sect. \ref{Grassmannrepresentation} we begin our discussion of the connection between a classical statistical ensemble for the two-level observables and fermionic Grassmann variables by a map between the local probability distributions and elements of a Grassmann algebra, and an associated map between two-level observables and Grassmann operators. Sect. \ref{Fermions} interprets the wave function $q_\tau$ as the wave function for a pure quantum state of a multi-fermion system. We recover the quantum  formalism with a Schr\"odinger equation for the wave function.

As a simple example, we describe in sect. \ref{Two-state} a classical two state system with a unitary evolution law for the classical probabilities. This time evolution is the same as for two state quantum mechanics for a spin in a constant magnetic field. We explicitly display the Grassmann representation for this system. In sect. \ref{Unequaltimecorrelation} we discuss within this example the issue of unequal time correlation functions and derivative observables as $\partial_t N$. We find that $\partial_tN(t)$ can be expressed in terms of the local probability distribution, but it involves an operator that does not commute with $N(t)$. Similarly, the expectation values of observables $N_\alpha(t+\Delta)$ can be computed from $q_\tau(t)$. They are represented by off-diagonal operators. We clarify how the emergence of non-commuting operators is related to the coarse graining of the information in the step from the all-time ensemble to the local-time subsystem. In sect. \ref{Measurements and quantum correlation} we address the connection between correlations of measurements and the non-commutative operator product. 

Sect. \ref{Quantumwave} discusses in more detail the wave function which is used to describe the unitary evolution.  A complex structure allows to map the real wave function $q_\tau(t)$ in eq. \eqref{A} to the usual complex wave functions in quantum mechanics. We give an example for a classical statistical ensemble leading to a complex wave function in sect. \ref{Twocomponentspinor}. It can describe a complex two-component spinor. In sect. \ref{Quantumfieldtheory} we address quantum field theories for fermions. We draw our conclusions in sect. \ref{Conclusions}. 

\section{All-time ensemble}
\label{all-time}
We consider a classical statistical ensemble of a generalized Ising type with $R$ bits. The states $\omega$ of this ensemble are bit sequences. At this stage no notion of time is introduced yet. We assume $R=TB$ and label the bits by $(n,\alpha)$, $n=1\dots T,~\alpha=1\dots B$. There are obviously many ordering structures of this type. The particular one that corresponds to time will be singled out by the existence of an evolution law that we discuss in sect. \ref{Evolutionlaw}. For the time being we simply assume that the appropriate selection is made and replace the label $n$ by a discrete time label $t$.

For a more general discussion of our setting we assume an ordered chain of points that we place at coordinates $t_i$ at equal distance, $t_{i+1}-t_i=\epsilon$. At each point we consider a set of numbers $n_\alpha(t_i)$ which can take the values one or zero. This may be interpreted as a property $\alpha$ realized at point $t_i~\big(n_\alpha(t_i)=1\big)$  or not $\big(n_\alpha(t_i)=0\big)$. A possible state $\omega$ of the system is then specified by an (ordered) sequence of numbers $n_\beta$ for all $t_i$,  $\omega=\big\{n_\beta(t))\big\}$. For only one property (no index $\alpha,\beta)$ it can be viewed as a chain of bits $\{n(t)\}$ taking values $0$ or $1$, one bit for every $t_i$. On the other hand, if we interpret $n_\beta(t_i)$ for a given $t_i$ itself as sequence, a state $\omega$ is a sequence of sequences. A classical statistical ensemble is specified by a probability $p_\omega\geq 0$ for every state $\omega$, with $\sum_\omega p_\omega=1$.

Classical observables $A$ have a fixed value $A_\omega$ in every state of the ensemble. Their expectation values obey
\be\label{1}
\kl A\kr=\sum_\omega A_\omega p_\omega.
\ee
We concentrate on the ``basis observables'' $N_\alpha(t)$ which can take the values $\big(N_\alpha(t)\big)_\omega=1,0$ according to
\be\label{2}
\big(N_\alpha(t)\big)_\omega\equiv N_\alpha(t)\big(\{n_\beta(t')\}\big)=n_\alpha(t).
\ee
In other words, $N_\alpha(t)$ ``reads out'' the bit $\alpha$ at a given point $t$ of the sequence of bits $\{n_\beta(t')\}$: for a given sequence it takes the value $1(0)$ if this sequence has for the particular values $\alpha$ and $t$ the number $n_\alpha(t)=1(0)$. Expectation values and classical correlations can be computed according to
\ba\label{3}
&&C^{(p)}_{\alpha_1,\alpha_2\dots\alpha_p}(t_1,t_2,\dots t_p)=
\kl N_{\alpha_1}(t_1)N_{\alpha_2}(t_2)\dots N_{\alpha_p}(t_p)\kr\nn\\
&&\hspace{1.5cm}=\sum_{\{n_\beta(t')\}}
p\big(\{n_\beta(t')\}\big)n_{\alpha_1}(t_1)\dots n_{\alpha_p}(t_p).
\ea
We can use the basis observables $N_\alpha(t)$ in order to construct ``composite observables'' by taking linear combinations or products of the basis observables. 

We assume a notion of ``completeness'' by postulating that all properties of reality are described by the probabilities $p_\omega$ and that no additional information is available.  We therefore deal with a genuinely probabilistic theory. We further assume in this paper that all physical observables can be constructed from the occupation numbers $N_\alpha(t)$. Then our setting is the most general classical statistical setting for states where the $N_\alpha(t)$ have fixed values. If there would be distinctions between two states $\omega_1,\omega_2$ which go beyond different values for some of the $N_\alpha(t)$, this would not affect the expectation values of physical observables. We can group $\omega_1$ and $\omega_2$ into a ``combined state'' $\omega$, with $p_\omega=p_{\omega_1}+p_{\omega_2}$. 

The probability distribution $\{p_\omega\}$ can be characterized by the classical correlation functions $C^{(p)}$ given by eq. \eqref{3}. Let us assume first a finite number of points, $i=1\dots T$, and a finite number of ``properties'' or ``species'', $\alpha=1\dots B$. We then have a total number $R=TB$ of ``bits'' $n_\alpha(t)$. The classical correlation functions can be linearly independent only for $p\leq R$. Indeed, $N_\alpha(t_i)$ is a projection operator
\be\label{4}
\big(N_\alpha(t_i)\big)^2=N_\alpha(t_i)
\ee
such that every factor $N_\alpha(t_i)$ can appear at most once in the independent correlations. The correlation with $p=R$ is unique, since it contains every $N_\alpha(t_i)$. There are $R$ correlation functions with $p=R-1$, which can be labeled by the $R$ possible places where one $N_\alpha(t)$ is missing from the chain of all $N_\beta(t_i)$. We can define the conjugate observable
\be\label{5}
\bar N_\alpha(t_i)=1-N_\alpha(t_i)
\ee
which takes the value $\big(\bar N_\alpha(t_i)\big)_\omega=1$ for all states $\omega$ where $\big(N_\alpha(t_i)\big)_\omega=0$ and $\big(\bar N_\alpha(t_i)\big)_\omega=0$ if $\big(N_\alpha(t_i)\big)_\omega=1$. One may then relate the correlation functions with $\bar p$ factors of $\bar N$ to the correlation functions with $p=R-\bar p$ factors of $N$. 

We define the manifold ${\cal P}$ of all possible probability distributions $\{p_\omega\}$ and ${\cal C}$ the manifold of all independent classical correlation functions. Eq. \eqref{3} constitutes a map ${\cal P}\to {\cal C}$. This map is injective since two probability distributions which yield the same value for all correlation functions are identical. Let us denote the image of ${\cal P}$ in ${\cal C}$ by ${\cal C}_p$. We can then find the inverse map ${\cal C}_p\to {\cal P}$. In other words, the probability distribution $\{p_\omega\}$ can be reconstructed from the classical correlation functions. 

We can easily visualize the situation in the simple cases of only one or two bits $(R=1,2)$. For $R=1$ we have only two states $n=0$ and $n=1$, with associated probabilities $p_-$ and $p_+$. The only ``correlation'' is the expectation value $(p=1)$
\be\label{6}
\kl N\kr=p_+-p_-.
\ee
With $p_++p_-=1$ this yields
\be\label{7}
p_\pm=\frac12(1\pm\kl N\kr).
\ee
For two bits $(n_1,n_2)$ we have four states, $\omega=\big[(1,1),(1,0),(0,1),(0,0)\big]$, with probabilities $\{p_\omega\}=[p_{++},p_{+-},p_{-+},p_{--}]$. The expectation values or ``occupation numbers'' obey
\be\label{8}
\kl N_1\kr=p_{++}+p_{+-}~,~\kl N_2\kr=p_{++}+p_{-+}.
\ee
There is one independent two-point correlation $(p=2)$ 
\be\label{9}
\kl N_1N_2\kr=p_{++}.
\ee
Together with $\sum_\omega p_\omega=1$ we can compute the probabilities $p_\omega$ from $\kl N_1\kr,\kl N_2\kr$ and $\kl N_1N_2\kr$. For three bits we have three expectation values $\kl N_1\kr, \kl N_2\kr,\kl N_3\kr$, three two point functions $\kl N_1N_2\kr,\kl N_1N_3\kr,\kl N_2N_3\kr$ and one three point function $\kl N_1N_2N_3\kr$. The values of the seven correlation functions are sufficient for the reconstruction of the seven independent probabilities $p_\omega$. This generalizes to an arbitrary number of bits $R$. We have $2^R$ different states $\omega$, $2^R-1$ independent probabilities $p_\omega$, and also $2^R-1$ independent correlation functions.

The classical correlation functions obtained from eq. \eqref{3} have to obey constraints - these constraints define ${\cal C}_p$. Obviously, all correlation functions obey the bounds
\be\label{10}
0\leq C^{(p)}_{\{\alpha_k\}}\big(\{t_k\}\big)\leq 1. 
\ee
There are further constraints. For the example of two bits the connected two-point-function obeys
\ba\label{11}
G^{(2)}&=&\kl N_1N_2\kr-\kl N_1\kr\kl N_2\kr\nn\\
&=&p_{++}p_{--}-p_{+-}p_{-+}~\in~
\left[-\frac14,\frac14\right].
\ea
This restriction  holds for the connected two-point-function also for an arbitrary number bits. If we single out two particular bits $n_1$ and $n_2$ out of the $R$ possible bits, we can compute the correlations \eqref{8}, \eqref{9} by ``integrating out'' the bits different from $n_1$ and $n_2$. This yields an effective probability $p_{++}$ for finding $N_1=1,N_2=1$, which is obtained by summing over all probabilities $p_\omega$ for those states for which the sequence $\{n_\alpha(t_i)\}$ has for $n_1$ and $n_2$ the value one. Similarly, we get $p_{+-},p_{-+}$ and $p_{--}$, and $\sum_\omega p_\omega=1$ translates to $p_{++}+p_{+-}+p_{-+}+p_{--}=1$. The relations \eqref{8} and \eqref{9} follow in a straightforward way. 

The map ${\cal P}\to {\cal C}_p$ can be used for a characterization of arbitrary statistical systems. If the classical correlation functions of the system obey constraints such they belong to ${\cal C}_p$ for an appropriate $R$, this system corresponds to a classical statistical ensemble with $R$ bits. For this ensemble, the probability distribution $\{p_\omega\}$ can be constructed from the full set of classical correlation functions. The constraints defining ${\cal C}_p$ may be used in order to test if a given statistical systems as a quantum system or Grassmann functional integral, can be represented by a classical $R$-bit system. More precisely, the test concerns both the existence of the classical ensemble and the use of classical correlation functions \eqref{3}. If correlation functions different from the classical correlations are used, the system may still be represented by a classical ensemble even if the test fails. This is important, for example, for the representation of quantum systems as classical statistical ensembles, where the correlation functions correspond to conditional correlations different from the classical ones \cite{QPSS}. 

The all-time ensemble is described by ``complete statistics'' in the sense that all classical correlation functions for all observables that are constructed from $N_\alpha(t)$ can be computed from the all-time probability distribution $\{p_\omega\}$. We will see later in this paper that this property is lost if we describe ``subsystems'' as the local-time subsystem discussed in the next section. Subsystems obtain by a ``coarse graining of the information''. Subsystem observables are those for which the expectation values can be computed from the information available for the subsystem. There is no guarantee that the classical correlation for two subsystem observables is itself a subsystem observable. We will discuss this issue which is very important for an understanding of quantum mechanics in more detail in sect. \ref{Unequaltimecorrelation}. 

\section{Local-Time subsystem}
\label{timesubsystem}

In this and the next section we introduce the notion of a local-time subsystem. For this purpose one infers from the all-time probability distribution $\{p_\omega\}$ a local-time probability distribution $\{p_\tau(t)\}$ for any given time $t$. A state of the local-time subsystem is given by the sequence of $\{p_\tau(t)\}$ for all times $t$. The local-time probability distribution $\{p_\tau(t)\}$ amounts to the usual definition of a statistical ensemble for which the probabilities relate to a particular time. The time evolution law for this ensemble will specify how $\{p_\tau(t+\epsilon)\}$ can be computed from $\{p_\tau(t)\}$. The local-time subsystem comprises all sequences $\{p_\tau(t)\}$ that obey the same evolution law. The different states of the local-time subsystem correspond to different sequences $\{p_\tau(t)\}$ that all realize the given evolution law.

\noindent
{\bf 1. \quad Locality in time}

So far our considerations hold for arbitrary bits without a distinction of the ``indices'' $\alpha$ and $t_i$. We could consider the pair $\gamma=(\alpha,t_i)$ as a ``collective index'' labeling all the $R$ bits. We next use the split between ``internal indices'' $\alpha$ and ``time indices'' $t_i$ in order to define local probabilities $p_\tau(t)$, where ``local'' means local in time. This is done by summing over all sequences $[n_\beta(t')]$ with $t'$ different from $t$,
\be\label{12}
p_\tau(t)=p\big([n_\beta(t)]\big);t)=\prod_{t'\neq t}\prod_\beta\sum_{n_\beta(t')=0,1}
p\big(\{n_\beta(t');n_\beta(t)\}\big).
\ee
The local probabilities depend on the sequence of bits $[n_\beta(t)]$ at a given time, where $\tau=[n_\beta(t)]$ can take ${\cal N}=2^B$ values. They are normalized
\be\label{12A}
\sum_\tau p_\tau(t)=\sum_{[n_\beta(t)]}
p\big([n_\beta(t)];t\big)=1.
\ee
The index $\tau=1\dots {\cal N}$ counts the number of ``local states'' at a given time $t$. In eq. \eqref{12} we have used a notation where $t$ singles out a particular time, while $t'$ extends over all other times, i.e.
\ba\label{13A}
\omega&=&\{n_\beta(t');n_\beta(t)\}=\big\{[n_\beta(t_1)],[n_\beta(t_2)],\dots,\nn\\
&&[n_\beta(t-\epsilon)],[n_\beta(t+\epsilon)]\dots [n_\beta(t_f)];[n_\beta(t)]\big\}.
\ea
For the local-time probability distribution we use the equivalent notations
\be\label{16AA}
p_\tau(t)\equiv p\big([n_\beta];t\big)\equiv p\big([n_\beta(t)];t\big)\equiv p\big([n_\beta(t)]\big).
\ee

The observables $N_\alpha(t)$ at a given time $t$ play the role of local observables. Their expectation values and ``equal-time correlation'' can be computed from $p_\tau(t)$, 
\ba\label{13}
\bar C^{(m)}_{\alpha_1,\dots,\alpha_m}(t)=\kl N_{\alpha_1}(t)\dots N_{\alpha_m}(t)\kr\nn\\
=\sum_{[n_\beta(t)]}p\big([n_\beta(t)];t\big)n_{\alpha_1}(t)\dots n_{\alpha_m}(t).
\ea
The general discussion for correlations is the same as above, except that all $N_\alpha$ are now taken at the same time $t$, and $R$ is replaced by $B,\alpha=1\dots B$. Integrating out the ``past degrees of freedom'' for $t'<t$ and the ``future degrees of freedom'' for $t'>t$ is formally straightforward, using the definition of the sums
\ba\label{14}
\sum_\omega&=&\sum_{\{n_\beta(t')\}}=\prod_{t'}\prod_\beta\sum_{n_\beta(t')=0,1}\nn\\
&=&\Big(\prod_\beta\sum_{n_\beta(t)=0,1}\Big)
\Big(\prod_{t'\neq t}\prod_\beta\sum_{n_\beta(t')=0,1}\Big)\nn\\
&=&\sum_{[n_\beta(t)]}\Big(\prod_{t'\neq t}\prod_\beta\sum_{n_\beta(t')=0,1}\Big).
\ea
Using eq. \eqref{12} we obtain eq. \eqref{13} from eq. \eqref{3}. 

We will also use the notion of ``extended locality'' by considering a time interval $t_a\leq \bar t\leq t_b$, with
\ba\label{15A}
&&\bar p\Big(\big\{n_\beta(\bar t)\big\}\Big)=\\
&&\prod_{t'<t_a} \prod_{t'>t_b} \sum_{n_\beta(t')}
p\Big(\big\{n_\beta(t'<t_a)~,~n_\beta(\bar t)~,~n_\beta(t'>t_b)\big\}\Big).\nn
\ea
The states $\big\{n_\beta(\bar t)\big\}$ are now sequences of $n_\beta$ for all allowed $\bar t$ within the interval. Correlation functions of observables $N_\alpha(t)$ at different times within the interval can be computed from the information contained in $\bar p\big(\big\{n_\beta(\bar t)\big\}\big)$.

\medskip\noindent
{\bf 2. \quad Periodic local probabilities}

The reader may get familiar with the concept of local probabilities by the discussion of a few simple examples. We discuss here periodic local probabilities, since they will play an important role in later parts of this paper. We concentrate first on only one species, $B=1$. Assume that the extended local probability density $\bar p\big(\big\{n(\bar t)\big\}\big)$ obeys the relation
\ba\label{15B}
&&\prod_{\tilde t\neq t_a,t_b}~~\sum_{n(\tilde t)=0,1}~\bar p\Big(\big\{1,n(\tilde t),0\big\}\Big)\nn\\
&=&\prod_{\tilde t\neq t_a,t_b}~~\sum_{n(\tilde t)=0,1}~
\bar p\Big(\big\{0,n(\tilde t),1\big\}\Big),
\ea
where the first and last entry in the sequence refer to the values of $n(t_a)$ and $n(t_b)$. In this case one finds for the local probability $p_1(t)=p\big(n(t)=1\big)$
\be\label{15C}
p_1(t_b)=p_1(t_a).
\ee
Indeed, the two expressions
\ba\label{AXA}
p_1(t_a)&=&\prod_{\tilde t\neq t_a,t_b}~~\sum_{n(\tilde t)=0,1} \Big(\bar p\big(1,n(\tilde t),1\big)+\bar p\big(1,n(\tilde t),0\big)\Big)\nn\\
p_1(t_b)&=&\prod_{\tilde t\neq t_a,t_b}~~\sum_{n(\tilde t)=0,1}~~
\Big(\bar p(1,\tilde n(\tilde t),1\big)+\bar p(0,n(\tilde t),1\Big),\nn\\
\ea
are equal by virtue of eq. \eqref{15B}. If the relation \eqref{15B} holds for all intervals with fixed size, $t_b-t_a=\tilde \tau$, one concludes that $p_1(t)$ is a periodic function of $t$ with period $\tilde \tau,p_1(t+\tilde\tau)=p_1(t)$. 

As a first example we take  a time chain with only four points $(T=4)$. We choose the probabilities
\ba\label{15D}
p(1,1,1,0)&=&p(1,0,1,1)=x,\\
p(1,1,1,1)&=&y-x~,~p(1,0,1,0)=1-y-x,\nn
\ea
and zero otherwise, with $0\leq y\leq 1~,~0\leq x\leq {\rm max }(y,1-y)$. This yields a period $2\epsilon$ for the local probabilities, with
\ba\label{15E}
p_1(t_1)&=&p_1(t_3)=1\nn\\
p_1(t_2)&=&p_1(t_4)=y.
\ea
For this example the four probabilities $p_1(t_i)$ constitute a state of the local-time subsystem. We can formally use periodic time by identifying $t_i+4n\epsilon=t_i~,~n\in N$. 

The number of constraints that have to be imposed on $p_\omega$ for obtaining a periodic local probability is typically far less than the number of independent $p_\omega$. For $T=4$ and period $2\epsilon$ the two equations of the type \eqref{15E} and the normalization of the probability distribution amount to three constraints for the $2^4$ probabilities $p(n_1,n_2,n_3,n_4)$. We conclude that many different probability distributions $\{p_\omega\}$ can lead to periodic local probabilities. The particular case \eqref{15E} imposes $p(0,n_2,n_3,n_4)=p(n_1,n_2,0,n_4)=0$ and therefore leaves only four nonvanishing probabilities. Then the probabilities \eqref{15D} are the most general solution of eq. \eqref{15E}. Still, a given local-time subsystem $p_\tau(t)$ \eqref{15E} is obtained for a whole family of all-time probabilities $p_\omega =p(n_1,n_2,n_3,n_4)$ that is parametrized by a continuous parameter $x$. The map $\{p_\omega\}\to \{p_\tau(t)\}$ is not invertible. A state of the local-time subsystem retains only part of the information that is available for the all-time probability distribution. 

A simple particular way of constructing all-time probability distributions leading to periodic local probabilities consists in ``gluing'' extended local probabilities. Consider extended local probabilities $\bar p\big(\big\{n_\beta(\bar t)\big\}\big)$ in the range $t_a\leq \bar t\leq t_b=t_a+\tilde\tau-\epsilon$. We may define an all-time probability distribution with $t$ in the range $t_a\leq t\leq t_a+m\tilde\tau-\epsilon$ by the product form
\ba\label{15F}
&&p\Big(\big\{n_\beta(\bar t)\big\}~,~\{n_\beta(\bar t+\tilde\tau)\big\}~,~\dots
\big\{ n_\beta\big (\bar t+(m-1)\tilde\tau\big)\big\}\Big)\\
&&=\bar p\Big(\big\{n_\beta(\bar t)\big\}\Big)\bar p
\Big(\big\{n_\beta(\bar t+\tilde\tau)\big\}\Big)\dots 
\bar p\Big(\big\{n_\beta(\bar t+(m-1)\tilde\tau\Big)\big\}.\nn
\ea
Using the normalization of the extended local probabilities $\bar p$ (for every fixed $k$, and with the product over $\bar t$ in the range of the corresponding extended local probability distribution),
\be\label{15G}
\prod_{\bar t+k\tilde\tau}~~\sum_{n_\beta(\bar t+k\tilde\tau)=0,1}
\bar p\Big(\big\{n_\beta(\bar t+k\tilde\tau)\big\}\Big)=1,
\ee
we obtain for the extended local probabilities in the interval $t_a+n\tilde\tau\leq\bar t+n\tilde\tau\leq\bar t+(n+1)\tilde\tau -\epsilon$ the distribution $\bar p\big(\big \{n_\beta(\bar t+n\tilde\tau)\big\}\big)$, which is the same for all $n$. 

The local probabilities are periodic with period $\tilde \tau$,
\be\label{28Aa}
p_\tau(t+n\tilde \tau)=p_\tau(t).
\ee 
Here we consider the states $\tau=[n_\beta]$ as given bit sequences, with $n_\beta$ corresponding to $n_\beta(t)$ or $n_\beta(t+n\tilde \tau)$ from the perspective of the all-time probability distribution. The periodicity follows from the general definition \eqref{12} of the local-time probability distribution, employing eq. \eqref{15F} for $\{p_\omega\}$. Time layers outside the extended time interval to which $t+n\tilde\tau$ belongs are integrated out trivially. The only remaining sums concern the times that belong to the same extended time interval as $t+n\tilde \tau$, but differ from this specific value, 
\ba\label{15H}
&&p\Big(\big [n_\beta(t+n\tilde\tau)\big ]\Big)\nn\\
&=&\prod_{\bar t+n\tilde\tau\neq t+n\tilde\tau}~~
\sum_{n_\beta(\bar t+n\tilde\tau)}
\bar p\Big(\big\{n_\beta(\bar t+n\tilde\tau)\big\},
\big[n_\beta(t+n\tilde\tau)\big]\Big).\nn\\
&=&p\Big(\big[n_\beta(t)\big]\Big).
\ea
The last line uses the fact that the extended local probability distributions $\bar p$ are the same for all $n$. 

As an example with $B=1~,~\tilde\tau=2\epsilon$, $t_a=t_1~,~t_b=t_2=t_1+\epsilon$ we may take 
\be\label{15I}
\bar p(1,1)=y~,~\bar p(1,0)=1-y~,~
\bar p(0,0)=\bar p(0,1)=0,
\ee
and consider the product \eqref{15F} for $m=2$
\ba\label{15J}
p(1,1,1,1)&=&y^2~,~p(1,0,1,0)=(1-y)^2,\nn\\
p(1,1,1,0)&=&p(1,0,1,1)=y(1-y)
\ea
(with vanishing $p$ otherwise). This is precisely the distribution \eqref{15D} for the special value $x=y(1-y)$.

Probability distributions $\{p_\omega\}$ leading to arbitrary periodic local probabilities $p_\tau(t)$ can be always constructed by gluing extended local probability distributions. These are, however, by far not the only possibilities. The all-time probability distribution \eqref{15J} is only one particular member of the family of probability distributions \eqref{15D} that lead to the same local-time subsystem \eqref{15E}. 

In this context we note that for glued all-time probability distributions the local-time probabilities $p_\tau(t)$ can be computed from the extended probability distribution in the time interval to which $t$ belongs. This holds for the whole sequence $\{p_\tau(t)\}$ for times belonging to this interval. Different extended probability distributions $\bar p$ can lead to the same sequence of local probabilities $\{p_\tau(t)\}$ within their associated time interval. An all-time probability distribution that realizes given periodic probabilities \eqref{28Aa} can therefore also be obtained by gluing different $\bar p$ that all lead to the same $\{p_\tau(t)\}$ within their time interval of definition.

We discuss next a further example with $B=1$ that we will use later in this paper. We consider an extended local distribution in the range $t_a\leq\bar t\leq t_a+3\epsilon$ for which the non-vanishing probabilities are given by
\be\label{15K}
\bar p(1,1,0,1)=\bar p(1,1,0,0)=\bar p(1,0,0,1)=
\bar p(1,0,0,0)=\frac14.
\ee
By gluing according to eq. \eqref{15F} we obtain a probability distribution which leads to a periodic local probability with period $\tilde\tau=4\epsilon$, namely
\be\label{15L}
p_1(t)=\cos^2\omega t~,~\omega=\frac{\pi}{\tilde\tau},
\ee
where $t=(n-1)\epsilon$. The state of the local-time subsystem is described by the sequence of local probabilities (with integer $k$)
\ba\label{30Aa}
p_1(k\tilde\tau)=1&,&p_1(k\tilde\tau+\epsilon)=\frac12,\nn\\
p_1(k\tilde\tau+2\epsilon)=0&,&p_1(k\tilde\tau+3\epsilon)=\frac12.
\ea
Again, there will be many different all-time probability distributions $\{p_\omega\}$ that lead to the same state of the local-time subsystem \eqref{30Aa}. 

We can extend this setting to a finer grid, $\tilde\tau=8\epsilon$, while keeping eq. \eqref{15L} unchanged with fixed $\tilde\tau$ and $\omega$. We have now to construct extended local probability distributions for eight time steps such that $p_1(j\epsilon)=\cos^2(j\pi/8), j=0\dots 7$. While the solution of the system of nine equations for the $2^8$ probabilities $\bar p\big(\big\{n(\bar t)\big\}\big)$ becomes cumbersome in practice, it is obvious that many such solutions exist. Different solutions can be glued together in order to obtain a local-time subsystem obeying eq. \eqref{15L}. Even these many possibilities are only a small subset of the all-time distributions $\{p_\omega\}$ that lead to the same state of the local-time subsystem \eqref{15L}. In a similar spirit we can construct probability distributions with a periodic local-time subsystem \eqref{15L} for arbitrary small values $\epsilon/\tilde\tau$. We can also take the limit $\epsilon\to 0$ for fixed $\tilde\tau$ and $\omega$. We will see in sect. \ref{Two-state} that that the simple state of the local-time subsystem \eqref{15L} describes two-state quantum mechanics.

\medskip\noindent
{\bf 3. \quad System observables}

The all-time probability distributions $\{p_\omega\}$ that lead to  a given state of the local-time subsystem \eqref{15L} define standard classical statistical ensembles. Let us now assume that only the information contained in the sequence of local probabilities $\{p_\tau(t)\}$ is available for predictions of the outcome of measurements. This is a familiar situation, since usually all quantities are computed from time-dependent probabilities, while an all-time probability distribution is not known. We therefore have to ask what type of quantities remain computable from the limited information contained in the state of the local-time system $\{p_\tau(t)\}$, e.g. the periodic probabilities \eqref{15L}. 

``System observables'' are those classical observables of the all-time ensemble for which the probabilities $w_a$ for finding a value $\lambda_a$ are computable in terms of the information available for the local-time subsystem. In particular, the expectation values remain computable in terms of the sequence of local probability distributions $\{p_\tau(t)\}$. This extends to expectation values of arbitrary functions of a system observable. If the spectrum of an observable contains only two values $\lambda_a$ it is sufficient that the expectation value is computable from $\{p_\tau(t)\}$. For observables $A$ with a spectrum containing more than two values $\lambda_a$ we have to require, in addition, that higher powers of $A$ have computable expectation values, i.e. $\kl A\kr, \kl A^2\kr,\dots$, etc. are computable from $\{p_\tau(t)\}$. 

The occupation numbers $N_\alpha(t)$ remain all system observables. This extends to the equal time correlation functions \eqref{13}. However, classical correlation functions involving occupation numbers at non-equal times are, in general, no longer computable from the information contained in the local-time subsystem. For example, the classical observable $N_\alpha(t_1)N_\beta(t_2)$ for $t_1\neq t_2$ is not a system observable.

One concludes that the local-time subsystem is characterized by incomplete statistics. While both $N_\alpha(t_1)$ and $N_\beta(t_2)$ are system observables, their classical product is not. This has important consequences for the definition of suitable correlations that describe a sequence of two measurements in the subsystem, as we will discuss in detail in sect. \ref{Unequaltimecorrelation}. If we take the attitude that only the information contained in the sequence of local probability distributions $\{p_\tau(t)\}$ for the local subsystem is available for an observer, the correlation between a first measurement of $N_\alpha$ at time $t_1$, and a second measurement of $N_\beta$ at time $t_2$, has to be computable in terms of $\{p_\tau(t)\}$. This implies that the appropriate ``measurement correlation'' cannot be given by the classical correlation function. 

\medskip\noindent
{\bf 4. \quad Equivalence classes for classical probabilities}

The sequence of local probabilities $p\big(\big[n_\beta(t_i)\big]\big)$ for all times $t_i$ can be  used for the definition of an equivalence class for the classical probability distribution  $\{p_\omega\}$. For any given $\{p_\omega\}$ the sequence of local probabilities can be computed in a unique  way by eq. \eqref{12}. Two probability distributions $\{p_\omega\}$ and $\{p'_\omega\}$ are  considered as equivalent if they yield the same local probabilities $p\big(\big[n_\beta(t_i)\big]\big)$  for all $t_i$. For a large number of time points $t_i$ the number $2^{TB}-1$ of independent real  $p_\omega$, which are needed to define the distribution $\{p_\omega\}$, exceeds by far the number $T(2^B-1)$ of independent real probabilities $p_\tau(t_i)$ needed for the specification of the equivalence class. Typically, a given equivalence class contains many different probability distributions $\{p_\omega\}$. We will see later that for many purposes the relevant properties of the system only depend on the equivalence classes of sequences of local probabilities. These equivalence classes correspond to the states of the local-time subsystem. 

We may also show that for each arbitrary sequence of local probabilities there exist classical probabilities $\{p_\omega\}$ that realize this sequence. (None of the equivalence classes is empty.) We proceed by induction, as we may first demonstrate for $B=1$. Assume that for $m$ time points of a sequence of local probabilities one has found a probability distribution $p^{(m)}\big(n(t_1)~,~n(t_2),\dots,n(t_m)\big)$ which yields all the wanted local probabilities $p_1(t_i)=p\big(n(t_i)=1\big)$. We can then construct
\ba\label{15M}
&&p^{(m+1)}\big(n(t_1),\dots n(t_m),1\big)=\nn\\
&&\hspace{1.5cm}p^{(m)}\big(n(t_1),\dots n(t_m)\big)p_1(t_{m+1}),\nn\\
&&p^{(m+1)}\big(n(t_1),\dots, n(t_m),0\big)=\\
&&\hspace{1.5cm}p^{(m)}\big((t_1),\dots,n(t_m)\big)
\big (1-p_1(t_{m+1})\big),\nn
\ea
in order to continue the sequence of local probabilities with an arbitrary $p_1(t_{m+1})$. For the first member $(m=1)$ of the induction chain we can take $p\big(n(t_1)=1\big)=p_1(t_1)$, $p\big(n(t_1)=0\big)=1-p_1(t_1)$. 

This  procedure can be generalized for arbitrary $B$. The first step starts with $p^{(1)}=p_1\big(\big[n_\beta(t_1)\big]$. In the second step the product 
\be\label{150}
p^{(2)}\big(\big[n_\beta(t_1)\big],\big[n_\beta(t_2)\big]\Big)=
p_1\Big(\big[n_\beta(t_1)\big]\Big)
p_2\Big(\big[n_\beta(t_2)\big]\Big)
\ee
realizes the sequence of local probabilities for $t_1$ and $t_2$, namely $p_1\big(\big[n_\beta)(t_1)\big]\big)~,~p_2\big(\big[n_\beta(t_2)\big]\big)$. This follows directly from the normalization \eqref{12A}. The third step multiplies $p^{(2)}$ by $p_3\big(\big[n_\beta(t_3)\big]\big)$ and so on. We recall that this particular construction only constitutes an existence proof that a suitable $\{p_\omega\}$ exists for every sequence of local probabilities. Many other and less trivial $\{p_\omega\}$ lead to the same sequence. 

\section{Evolution law}
\label{Evolutionlaw}
We next postulate as a key ingredient for a time structure that the local probabilities $p_\tau(t)$ can be expressed in terms of the neighboring local probabilities $p_\tau(t-\epsilon)$. This relation will be called the ``law of evolution''. The all-time probability distribution $p_\omega=p\big(\{n_\beta(t)\}\big)$ contains statistical information for arbitrary times, past, present and future. The existence of a law of evolution is a restriction on the allowed $\{p_\omega\}$. The law of evolution allows us, in principle, to compute correlation functions at some ``present time'' $t$ from probability distributions in the past without using any information concerning the future. It is sufficient to know the probability distribution at some ``initial time'' $t_0$, and to solve the law of evolution in order to compute $p_\tau(t')$ for $t_0<t'\leq t$. Most important, the evolution law allows us to make probabilistic predictions for future expectation values and equal time correlations with time arguments $t_k>t$, using only statistical information from the present or past correlations. Also the (extended) local correlation functions can be determined using much less information than contained in the ``overall probability distribution '' $\{p_\omega\}$. If we are interested only in times $t_k>t_0$ the existence of an evolution law is only needed for $t\geq t_0$. 

\medskip\noindent
{\bf 1. \quad Simple periodic evolution}

As an example we consider $B=1$  and the periodic state \eqref{15L} of the local-time subsystem given by $p_1(t)=\cos^2\omega t$. One can obviously write down a differential evolution law to which $p_1(t)$ is a solution, as $\partial^2_tp_1=-\omega^2p_1$. This involves, however, a second time derivative such that the computation of $p_1(t)$ requires knowledge of $p_1(t-\epsilon)$ and $p_1(t-2\epsilon)$. We are interested in an evolution law that involves only one time derivative, such that $p_1(t)=p_1(t-\epsilon)+\epsilon\partial_t p_1(t_1-\epsilon)$ is computable from $p_1(t-\epsilon)$ if $\partial_tp_1$ can be expressed in terms of $p_1$. Such an evolution law will no longer be linear in $p_1$, since 
\be\label{36A}
\partial_t p_1=-2\omega\cos \omega t\sin \omega t=-2\omega s\sqrt{p_1(1-p_1)}.
\ee
The sign $s=\pm 1$ depends on $t$, with $s=1$ for $t$ in the interval $[0,\pi/2\omega]$ and $s=-1$ for $t\in [\pi/2\omega,\pi/\omega]$, and periodically continued with period $\tilde \tau=\pi/\omega$. The sign jumps whenever $p_1$ or $1-p_1$ equal zero, resulting in $\partial_tp_1=0$. We need this prescription for the sign changes of $s$ in order to obtain the correct evolution law. The computation of $\partial_tp_1(t)$ not only needs $p_1(t)$, but also the information if $s$ is positive or negative for a given time $t$. 

A convenient form to combine the information in $p_1(t)$ and $s(t)$ is the two-component real wave function 
\ba\label{36B}
q_1(t)&=&\bar s(t)\sqrt{p_1(t)},\nn\\
q_0(t)&=&\bar s(t)s(t)\sqrt{1-p_1(t)},
\ea
which is normalized according to 
\be\label{36C}
q^2_1(t)+q^2_0(t)=1.
\ee
The evolution equation 
\be\label{36D}
\partial_tq_1=-\omega q_0~,~\partial_tq_0=\omega q_1
\ee
describes a simple rotation among the two components of a unit vector. The overall sign $\bar s$ in eq. \eqref{150} does not affect the evolution equation \eqref{36D}, but the relative sign $s$ between $q_1$ and $q_0$ matters. (For definiteness we may take $\bar s=1$ for $t$ in the interval $[-\pi/2\omega, \pi/2\omega]~mod~ \tilde \tau$, and $\bar s=-1$ otherwise.) 

The evolution equation \eqref{36D} is linear and avoids the explicit appearance of a sign factor. It is straightforward to compute $q_1(t)$ and $q_0(t)$ from eq. \eqref{36D} if $q_1(t-\epsilon)$ and $q_0(t-\epsilon)$ are known. A first order evolution equation is obviously formulated much easier in terms of the real wave function $\big (q_1(t),q_0(t)\big)$ than in terms of the probabilities. We will argue next that evolution equations of the type \eqref{36A} or \eqref{36D} arise naturally in the context of periodic probability distributions. We will do so by showing first that first order evolution equations which are linear in the probability distribution $p(t)$ usually do not lead to periodic probabilities.

\medskip\noindent
{\bf 2. \quad Transition matrix}

For a general discussion of possible evolution laws we first discuss a ''linear law of evolution'' which can be written in terms of a ``transition matrix'' $W$
\be\label{15}
p_\tau(t)=\sum_\rho W_{\tau\rho}(t,t-\epsilon)p_\rho(t-\epsilon).
\ee
If $W$ obeys 
\be\label{16}
\sum_\tau W_{\tau\rho}=1~,~\sum_\rho W_{\tau\rho}=1
\ee
and
\be\label{17}
W_{\tau\rho}\geq 0,
\ee
the matrix element $W_{\tau\rho}$ can be interpreted as the conditional probability to find the state $\tau$ at time $t$ if the state $\rho$ is realized at time $t-\epsilon$. Eq. \eqref{17} guarantees that conditional probabilities (or transition probabilities from $\rho$ to $\tau$) are positive, and the first eq. \eqref{16} ensures the correct normalization, namely that one of the states $\tau$ must be found at $t$ if $\rho$ is realized at $t-\epsilon$. Eq. \eqref{15} states that the probability to find at $t$ the state $\tau$ equals the sum over all states $\rho$ at $t-\epsilon$, weighted with their respective probabilities and the transition probability from $\rho$ to $\tau$. The second eq. \eqref{16} states that every state $\tau$ at $t$ origins from one of the states $\rho$ at $t-\epsilon$ with probability $W_{\tau\rho}$, and that these probabilities sum up to one. Even though the linear evolution law \eqref{15}-\eqref{17} is very intuitive we will see that it cannot describe the oscillating probabilities that are omnipresent in nature. 

We discuss in some more detail in appendix A that one of the main reasons for this shortcoming is the lack of time reversal symmetry. Time reversal symmetry can be realized in a straightforward way in the class of ``unitary evolution laws'' that we will discuss next.

\medskip\noindent
{\bf 3. \quad Unitary time evolution}

If we want to find a general description of oscillating  probabilities we have to look for an alternative to the linear evolution law \eqref{15}-\eqref{17}. Indeed, there is no need for the relation between $\{p_\tau(t)\}$ and $\{p_\rho(t_0)\}$ to be linear as we have seen for the simple example \eqref{36A}.  An interesting alternative to the transition matrix \eqref{21Aneu} is a time evolution of the local probabilities according to
\be\label{21A}
p_\tau(t)=\sum_{\rho,\sigma}V_{\tau\rho\sigma}(t,t_0)
\sqrt{p_\rho(t_0)p_\sigma(t_0)},
\ee
where $V_{\tau\rho\sigma}$  obeys
\be\label{21AA}
\sum_\tau V_{\tau\rho\sigma}(t,t_0)=\delta_{\rho\sigma}.
\ee
This can be cast into the form of an evolution law by taking $t_0=t-\epsilon$. It is the type of evolution law obeyed by the periodic local probability \eqref{15L}. For only one species $(B=1)$ the indices $\tau,\rho,\sigma$ take values $(1,0)$ with $p_0=1-p_1$. It is straightforward to verify that the evolution law \eqref{21A} with $V_{\tau\rho\sigma}(t,t-\epsilon)$ given by
\ba\label{26A}
V_{111}&=&V_{000}=\cos^2(\omega\epsilon)~,~V_{100}=V_{011}=\sin^2(\omega\epsilon),\\
V_{110}&=&V_{101}=-V_{010}=-V_{001}=-s(t-\epsilon)\cos(\omega\epsilon)\sin(\omega\epsilon),\nn
\ea
is obeyed by the local probability \eqref{15L}. The sign $s(t-\epsilon)=\pm 1$ is positive for the first half period, $0\leq t-\epsilon<\tilde\tau/2$ and negative for the second, $\tilde\tau/2\leq t-\epsilon <\tilde\tau$, with periodic continuation. 

The non-linear evolution law \eqref{21A} corresponds to a simple linear evolution law for the classical wave function, which amounts to the root of the probability distribution up to signs. Indeed, simple evolution laws are most easily formulated in terms of the classical wave function $\{q_\tau\}$. For an evolution that preserves the norm of the probability distribution it is rather natural to consider a unit vector $\vec q$ with components
\be\label{21B}
q_\tau=s_\tau \sqrt{p_\tau}~,~s_\tau=\pm 1.
\ee
The conserved normalization $\sum_\tau p_\tau=1$ corresponds to the unit  length of 
$\vec q,~\vec q^2=\sum\nolimits_\tau q^2_\tau=\sum\nolimits_\tau p_\tau=1$. By construction the inequality $0\leq p_\tau\leq 1$ is guaranteed for arbitrary $q_\tau$. Any time evolution that preserves the normalization of the probability distribution must keep the length of the vector $\vec q$ fixed and therefore corresponds to a rotation of $\vec q$, 
\ba\label{21C}
q_\tau(t)&=&\sum_\rho R_{\tau\rho}(t,t_0)q_\rho(t_0),\nn\\
\vec q(t)&=&R(t,t_0)\vec q(t_0)~,~
RR^T=1.
\ea

The simplest form of an evolution law is linear in $\vec q$, i.e. the rotation matrix $R$ does not depend on $\vec q$. This is realized by the simple periodic probability distribution \eqref{15L} discussed in sect. \ref{timesubsystem}. We will assume in the following that $R$ is independent of $\vec q$. Linearity in $\vec q$ entails, however, non-linearity in $\{p_\tau\}$ of the specific form of eq. \eqref{21A}, where the coefficients $V_{\tau\rho\sigma}(t,t_0)$ can be related to the rotation matrix $R_{\tau \rho}(t,t_0)$ up to a sign
\be\label{21D}
V_{\tau\rho\sigma}(t,t_0)=R_{\tau\rho}(t,t_0)R_{\tau\sigma}(t,t_0)
~\text{sign}~\big(q_\rho(t_0)q_\sigma(t_0)\big) .
\ee
Eq. \eqref{21AA} follows then from $R^TR=1$. For the example \eqref{15L}, \eqref{26A} one has for $R(t,t-\epsilon)$ 
\be\label{29A}
R_{11}=R_{00}=\cos(\omega\epsilon)~,~R_{01}=-R_{10}=\sin(\omega\epsilon).
\ee
The two-component vector $q_\tau(t)$ is given by $q_1(t)=\cos(\omega t)$, $q_0(t)=\sin(\omega t)$. 

We will see later that many quantum features can be related to the evolution law \eqref{21C} which is linear in the classical wave function $\vec q$. It is therefore worthwhile to point out that no ``quantum features'' are required or ``put in'' in the formulation of eq. \eqref{21C}. It is simply the most straightforward way to realize a norm-conserving time evolution of $\{p_\tau(t)\}$ and to realize oscillating probability distributions. We also emphasize that the classical wave function is real and that the phases characteristic for quantum mechanics have to emerge at a later stage, cf. sect. \ref{Twocomponentspinor}.

The rotations \eqref{21C} lead to a simple composition law, now in terms of a matrix multiplication of the rotation matrices
\be\label{28A}
R(t_3,t_1)=R(t_3,t_2)R(t_2,t_1).
\ee
An inverse rotation matrix always exists such that a time reflected evolution can be realized. Probability distributions $p(\{n_\beta(t_i)\})$, which obey the evolution law \eqref{21A}, \eqref{21C}, \eqref{21D}, have all the properties necessary for consistently determining the future from the past in a probabilistic way, as described in the beginning of this section.

We will call the evolution law \eqref{21C} a ``unitary evolution'', since rotations are a special case of the more general unitary transformations. In the remainder of this paper we will concentrate on unitary evolution laws. Obviously, a unitary time evolution is well suited for the description of oscillating probabilities. It is sufficient to use periodic rotations. The infinitesimal form of the unitary evolution law \eqref{21C} reads
\be\label{56A}
\partial_t q=Kq~,~K=(\partial_tR)R^{-1}.
\ee
The matrix $K$ is antisymmetric. 

\medskip\noindent 
{\bf 4. \quad States of local-time subsystem}

The local-time subsystem is characterized by the sequences of local-time probabilities $\{p_\tau(t)\}$ that obey a specific evolution law. A state of the local-time subsystem corresponds to a particular sequence. Formally, each state involves the information contained in many local probability distributions $\{p_\tau(t)\}$, one for each time. 

The presence of an evolution law greatly reduces the information needed for the specification of a state. the reason is the restriction on the allowed $\{p_\tau(t)\}$ imposed by the evolution law. For an evolution law formulated as a first order differential equation the allowed $\{p_\tau(t)\}$ must be solutions of this equation. As an example we may consider the local-time subsystem which is defined by the evolution equation \eqref{36A}. The necessary information for the specification of a state is the probability $p_1(t_{in})$ for some initial time $t_{in}$, the sign $s(t_{in})$, and the rule that the sign switches whenever $\partial_t p_1=0$. This is somewhat cumbersome due to the explicit appearance of the sign function $s(t)$. 

For a unitary time evolution law the specification of a state becomes much easier by use of the classical wave function $\{q_\tau(t)\}$. Indeed, any particular solution of the linear differential equation \eqref{36D} or \eqref{21C}  is uniquely determined by the wave function $\{q_\tau(t_{in})\}$ at some arbitrary initial time $t_{in}$. In short, the local-time subsystem is defined by the matrix $K$ in eq. \eqref{56A}, and a particular state of the local-time subsystem can be characterized by the classical wave function $\{q_\tau(t)\}$. In turn $\{q_\tau(t)\}$ is uniquely determined by an initial value $\{q_\tau(t_{in})\}$. 

\medskip\noindent
{\bf \quad  5. Equivalence classes for classical observables}

The system observables define equivalence classes for classical observables. Two different classical observables $A$ and $A'$, with different values $A_\omega$ and $A'_\omega$ in the states of the all-time ensemble, are considered as equivalent if they are both system observables and if they have the same probabilities $w_a$ for finding a value $\lambda_a$ for all states of the local-time subsystem. Two equivalent observables cannot be distinguished by any measurement of properties of the local-time subsystem. From the point of view of the subsystem they are identical. Two-level observables have only two possible values for $\lambda_a$. They are equivalent if their expectation value is the same for all arbitrary sequences $\{p_\tau(t)\}$ that obey the evolution law which specifies the local-time subsystem.

A simple example may be given for the local-time subsystem with evolution equation \eqref{36D}, corresponding to $K=-i\omega \tau_2$. All allowed wave functions are periodic in time, with period $2\pi/\omega$. Furthermore, after a period $\tilde\tau=\pi/\omega$ the wave function has only changed its overall sign such that the local-time probability distribution $\{p_\tau(t)\}$ is periodic with period $\tilde\tau$. In consequence, the occupation numbers $N(t)$ and $N(t+\tilde\tau)$ have the same expectation values for all states of the local-time subsystem. They belong to the same equivalence class.

The explicit classification which classical observables belong to the same equivalence class may be quite involved, in particular for more complicated evolution laws. The mathematical reason for the existence of equivalence classes is quite simple, however. We may restrict the discussion to two-level observables for simplicity. Two classical two-level observables $A$ and $A'$ are different if there exists at least one probability distribution $p_\omega$ for which $\kl A\kr\neq \kl A'\kr$. Let us denote by $\cP_1$ the ensemble of $\{p_\omega\}$ for which $\kl A\kr\neq \kl A'\kr$, and by $\cP_2$ the ensemble for which $\kl A\kr=\kl A'\kr$. The two observables $A$ and $A'$ are equivalent if all probability distributions $\{p_\omega\}$ that obey the given evolution law belong to $\cP_2$. 

More precisely, we may denote by $\cP_l$ the ensemble of all-time probability distributions $\{p_\omega\}$ that obey the given evolution law. This is typically only a small subset of the ensemble of all possible probability distributions $\{p_\omega\}$. The complement of $\{p_\omega\}$ not obeying the evolution law is denoted by $\bar \cP_l$. There are many pairs of classical observables $A$ and $A'$ such that $\cP_l \subset\cP_2$. Such observables have different expectation values only for distributions $p_\omega$ that belong to $\bar \cP_l$. An observable $A$ is a system observable if it has the same expectation value $\kl A\kr$ for all $p\in \cP_l$ that lead to the same state of the local-time subsystem. In other words, $\kl A\kr$ depends only on the equivalence class of classical probabilities that is specified by a given sequence of local-time probabilities $\{p_\tau(t)\}$. Consider now a pair of two observables for which $P_l \subset P_2$. If one is a system observable the other is as well. The two observables are equivalent. In other words, two equivalent classical two-level observables are distinguished only by different expectation values for some $p_\omega\in \bar\cP_l$. 

In turn, we may characterize a system observable by the probabilities $w_a\big(\{p_\tau(t)\}\big)$ for the possible values $\lambda_a$ for all states of the local-time subsystem. This is sufficient from the point of view of the local-time subsystem. However, no unique classical observable is singled out by this prescription which fixes only the equivalence class. We will see that this situation has important consequences for an understanding of quantum mechanics. The system observables will be associated with quantum operators. The lack of an unique classical observable associated to a quantum operator violates a basic assumption of the Kochen-Specker theorem \cite{KS}. This is why this theorem does not apply to a realization of quantum mechanics by a classical statistical ensemble \cite{QPSS}. 

\section{Quantum formalism}
\label{Quantum formalism}

Our discussion has proceeded purely in the context of classical statistics. The all-time probability distribution $\{p_\omega\}$ defines a standard classical statistical ensemble, and the observables are standard classical observables. As we have argued in ref. \cite{CW1} it can be convenient to use the quantum formalism also for classical statistical ensembles. This is based on the ``classical wave function'' which amounts to the root of the probability distribution up yo a sign. It is precisely this wave function that appears naturally in the formulation  of a unitary time evolution law. The quantum formalism arises therefore quite naturally for the description of the local-time subsystem with unitary evolution law. 

\medskip\noindent
{\bf 1. \quad Probability amplitude and Schr\"odinger 

\hspace{0.55cm}equation}

Our discussion of oscillating local probabilities has led us to the concept of a ``probability amplitude'' $q_\tau(t)$. In the limit $\epsilon\to 0$ we may cast the rotation of $\vec q$ in eq. \eqref{21C}, \eqref{56A} into the form of a differential ``Schr\"odinger equation''
\ba\label{30A}
\partial_t q&=&-iHq,\nn\\
H&=&iK=(\partial_tR)R^{-1}=\sum_z\alpha_zL_z.
\ea
Here $L_z$ are the $2^{B-1}(2^B-1)$ hermitean generators of the orthogonal group $SO({\cal N})=SO(2^B)$, which are represented here as purely imaginary antisymmetric matrices, normalized to have eigenvalues $0$ or $\pm 1$, and $\alpha_z\in {\mathbbm R}$. At this level the wave function $q(t)$ is real for all $t$. We can cast the Schr\"odinger equation into the more usual form with a ${\cal N}/2$-component complex wave function $\psi(t)$ whenever $H$ is a linear combination of generators belonging to the $U({\cal N}/2)$ subgroup of $SO({\cal N})$. (For time-independent $\alpha_z$ this can always be realized formally. However, a useful complex representation also requires that the observables which are relevant for the system are compatible with the complex structure.) On the other hand, a complex quantum mechanical wave function can always be written as a real wave function with twice the number of components. 

The rotations $R$ can also be expressed in terms of elementary bit operators, now acting on $q$
\be\label{30B}
R=\sum_v\rho_vB^{(v)}.
\ee
The relation $R^TR=1$ restricts the allowed values for the real constants $\rho_v$. An example is $R=R^T=B^{(1)}_-+B^{(1)}_+,R^2=1$, cf. eq. \eqref{21}. This combined transformation is a bit operator which flips the bit $1,(1,n_2)\leftrightarrow(0,n_2)$. Similarly, we can also express the Hamilton operator \eqref{30A} or the $SO({\cal N})$ generators $L_z$ in terms of bit operators (with imaginary coefficients). For the special case of a matrix $R_{\tau\rho}$ which has in every line only one nonzero entry one has $R_{\tau\rho}R_{\tau\sigma}=W_{\tau\rho}\delta_{\rho\sigma}=V_{\tau\rho\sigma}~,~W_{\tau\rho}=
(R_{\tau\rho})^2\geq 0~,~\sum_\tau W_{\tau\rho}=1$ and $p_\tau(t)=\sum_\rho W_{\tau\rho}p_\rho(t_0$). For this special situation the unitary evolution law \eqref{21C} results in a linear evolution law for the probabilities.

\medskip\noindent
{\bf 2. \quad Maps to quantum mechanics}

For probability distributions that obey a unitary evolution law one can construct maps to quantum systems. This allows to use the quantum formalism for establishing results for classical probability distributions, as we will see in the discussion of equivalence classes below. Typically, many different maps to quantum systems exist and they are often rather formal and will be suited only for the demonstration of general properties. We will see in later sections, however, that also realistic quantum systems as quantum field theories for fermions can be obtained in this way.

The simplest map uses the real wave function $\vec q$ with a purely imaginary Hamiltonian given by eq. \eqref{30A}. For a purely imaginary Hamiltonian a real wave function is no conceptional restriction, since the time evolution would keep the real and imaginary components of a complex wave function separated. All classical observables based on the occupation numbers $N_\alpha(t)$ at a given time are represented by real diagonal quantum observables $\big (\hat N_\alpha(t)\big)_{\tau\tau'}=\big(N_\alpha(t)\big)_\tau\delta_{\tau\tau'}$. Their expectation values can be computed by the usual quantum rule 
\ba\label{52A}
\kl N_\alpha(t)\kr&=&\kl q|\hat N_\alpha(t)|q\kr\nn\\
&=&\sum_{\tau,\tau'}q_\tau(t)\big(\hat N_\alpha(t)\big)_{\tau\tau'}q_{\tau'}(t).
\ea

A second map uses the analogy between Ising spins and occupation numbers for fermions. Since the ``local states'' are characterized by a sequence of occupation numbers $n_\alpha(t)=(0,1)$ we can associate a given state $\tau=\big [n_\alpha(t)\big]$ with a basis state for a multi-fermion system in the occupation number basis. Fermions with $B-1$ ``one particle states'' can be described in quantum mechanics by a $2^{B-1}$-component complex vector in the occupation number basis. This is equivalent to a $2^B$-component real vector and we take the last bit for a distinction between the real part of a complex wave function for $n_B=1$, and the imaginary part for $n_B=0$. (A quantum  time evolution with a real hermitean Hamiltonian amounts then to oscillations between states with $n_B=1$ and $n_B=0$.) 

In the fermionic quantum system the observables are usually written in terms of annihilation and creation operators $a_k,a^\dagger_k,k=1\dots B-1$. With $\alpha=(k,B)$ they can be directly transferred to our system of two-level observables. In particular, the fermionic occupation numbers $N_k=a^\dagger_k a_k$ are represented by diagonal operators. The eigenvalues $0,1$ correspond directly to the spectrum of the classical observables $N_\alpha(t)$ for $\alpha=k$. The total number of fermions in the quantum system corresponds to $N_F=\sum_kN_k$. We may split $k=(\vec x,s)$ into a space index $\vec x$ and an internal index $s$, with $\vec x$ a vector on a $d$-dimensional lattice with lattice distance $\eta$ and volume $V$. In the limit $\eta\to 0,V\to\infty$, corresponding to $B\to\infty$, our system describes a quantum type of field theory for fermions in $d$-dimensions, with internal degrees of freedom labeled by $s$. 

The map to systems has been constructed only on the level of the local-time probabilities. The time evolution of a ``many body wave function'' 
\ba\label{52A}
&&\psi\Big(\big[n_s(\vec x)\big];t\Big)=\\
&&q\Big(\big[n_s(\vec x)\big],n_B=1;t\Big)+iq\Big(\big[n_s
(\vec x)\big],n_B=0;t\big)\nn
\ea
is given by a Schr\"odinger equation. The translation of the Hamiltonian \eqref{30A} to the complex basis yields a hermitean Hamiltonian which has, in general, both real and imaginary parts.

\medskip\noindent
{\bf 3.\quad Sequences of local probabilities }

For the discussion of this section we do not need the explicit form of the ``overall probability distribution'' $\{p_\omega\}$. It is sufficient that such a distribution exists, as we have established in sect. \ref{timesubsystem}. Indeed, the evolution law together with the ``initial condition'' $q_\tau(t_0)$ at some arbitrary $t_0$ is sufficient in order to fix the whole sequence of local  probabilities $p_\tau(t_i)$ for all $t_i$. (For the unitary evolution this holds automatically for $t_i>t_0$ and $t_i<t_0$, since $R$ is invertible. For an evolution according to a transition matrix the computation of $p_\tau(t_i<t_0)$ needs the invertibility of $W$. Otherwise $t_0$ should be chosen as the smallest time.) The evolution law and the initial data therefore fix the equivalence class discussed in sect. \ref{timesubsystem} uniquely. For a unitary evolution we may therefore label the state of the local-time subsystem by the evolution law (the matrix $K$) and the initial condition for $q_\tau(t_0)$. (Here we assume some particular choice of signs for $q_\tau(t_0)$. If two different $q_\tau(t_0)$ and $q'_\tau(t_0)$ lead to the same sequence of probabilities $p_\tau(t)$ for all $t$ they actually belong to the same equivalence class. This holds if $q_\tau(t_0)$ and $q'_\tau(t_0)$ cannot be distinguished by the relevant observables. Otherwise we can refine the notion of equivalence classes by basing them on sequences $q_\tau(t)$.) 

We want to know which expectation values and correlation functions can be computed in terms of the sequences of local probabilities $p_\tau(t)$, without invoking further information from the all-time probability distribution $\{p_\omega\}$. For this type of question the mapping to the quantum formalism is very useful. In quantum mechanics we know that the Hamiltonian and the initial value of the wave function $q_\tau(t_0)$ are sufficient for the computation of expectation values of observables and their quantum  correlations at different times. All these properties depend therefore only  on the equivalence class of a sequence of local probabilities, and not on the detailed all-time probability distribution $\{p_\omega\}$. Most of the information contained in $\{p_\omega\}$ is not necessary for the expectation values and quantum correlations. One may argue that for many circumstances the physical properties of a system can be characterized by the quantum correlations. In this case the knowledge of the equivalence classes given by $q_\tau(t)$ is sufficient. In the other direction one may postulate that the outcome of measurements should only depend on the equivalence classes of sequences of probability distributions. Then the quantum correlations are good candidates for a description of the outcome of a sequence of measurements.

We emphasize in this context that the quantum correlations are based on conditional probabilities as discussed in detail in \cite{QPSS}. They are different from the ``classical correlations'' defined by eq. \eqref{3}. Indeed, two different probability distributions $\{p_\omega\}$ and $\{p'_\omega\}$ which belong to the same equivalence class typically lead to different classical correlations. If only the information concerning the equivalence class of sequences of local probabilities is available, the classical correlations are often not computable. On the level of the local probabilities we deal then with ``incomplete  statistics'' \cite{3,QPSS}.

\section{Grassmann representation}
\label{Grassmannrepresentation}

The analogy with fermions motivates the formulation of our Ising-type classical statistical ensemble with states characterized by the values of two-level observables in terms of Grassmann variables. This leads to simple expressions for the bit-operators. Besides the technical advantages of the Grassmann formulation the next sections will show how the basic concepts of quantum field theory, namely fermions, quantum states and unitary time evolution, arise in a natural way from our setting of probabilistic time. Once fermions are implemented, bosons can easily be constructed as suitable composite fields for fermion bilinears or more complicated bosonic composite fields. This may include gravity and gauge interactions as in spinor gravity \cite{HC}.

\medskip\noindent
{\bf 1. \quad Grassmann algebra}

In this section we construct the representation of states, probabilities, observables and evolution law in terms of a Grassmann algebra. We restrict our discussion here to the ``local probability distribution'' $p\big([n_\beta(t)]\big)$ at a given time $t$. To each bit we associate a Grassmann variable $\psi_\alpha,\alpha=1\dots B$. Grassmann variables anticommute and obey the standard differentiation and integration rules
\ba\label{22}
\psi_\alpha\psi_\beta+\psi_\beta\psi_\alpha&=&0,\\
\frac{\partial}{\partial\psi_\beta}\psi_\beta \psi_{\alpha_1}\dots\psi_{\alpha n}&=&
\int d\psi_\beta\psi_\beta\psi_{\alpha_1}\dots \psi_{\alpha_n}=
\psi_{\alpha_1}\dots\psi_{\alpha_n}.\nn
\ea
To every state $\tau=[n_\beta]$ we associate a basis element of the Grassmann algebra, $\tau\to g_\tau$, which is a product of factors $\psi_\beta$ according to the following rule: if within the sequence $[n_\beta]$ one has for a given $\alpha$ the value $n_\alpha=0$, we take a factor $\psi_\alpha$ in $g_\tau$, while $n_\alpha=1$ corresponds to a factor $1$. The order of the factors $\psi_\alpha$ corresponds to the order in the bit sequence. Thus an ordered chain $(n_1,n_2,\dots,n_B)$ corresponds to an ordered product of factors $\psi_\alpha$ for each $n_\alpha=0$ in the chain. An example for the map from the bit chain to the Grassmann algebra $\tau\to g_\tau$ is 
\be\label{23}
(1,0,0,1,0,\dots)\to \psi_2\psi_3\psi_5\dots
\ee
Since the order of the Grassmann variables in a product matters, we have to fix for this purpose some arbitrary ordering of the bits such that they are labeled in order by $\alpha=(1,2,\dots B)$. The sign convention for the basis elements $g_\tau$ has a plus sign if all factors $\psi_\alpha$ are ordered such that smaller $\alpha$ are to the left. The Grassmann elements $g_\tau$ form a complete basis in the sense that an arbitrary element $g$ of the real Grassmann algebra can be written as $g=\sum_\tau\lambda_\tau g_\tau~,~\lambda_\tau\in {\mathbbm R}$. In other words, an element of the Grassmann algebra can be specified by the set of real numbers $\{\lambda_\tau\}$. 

As a next step, we associate to an arbitrary probability distribution $\{p_\tau\}$ an element of the Grassmann algebra $g\in {\cal G}$. Since this association holds only up to a sign it is convenient to use directly the isomorphism between the real classical wave functions $\{q_\tau\}$ and the elements $g$ of a real Grassmann algebra,
\be\label{24}
g=\sum_\tau q_\tau g_\tau~,~p_\tau=q^2_\tau.
\ee
This defines a map ${\cal G}\to {\cal P}~,~g\to \{p_\tau\}$ realized by $g\leftrightarrow \{q_\tau\},\{q_\tau\}\to\{p_\tau\}$. The map is not invertible since the components $q_\tau$ are fixed by $p_\tau$ only up to a sign. Only for a given sign convention $q_\tau=s_\tau\sqrt{p_\tau}$, $s_\tau=\pm 1$, one can define an inverse map ${\cal P}\to{\cal G}$. 

We also define the conjugate element $\tilde g$ of $g$ by
\be\label{25}
\tilde g=\sum_\tau q_\tau \tilde g_\tau,
\ee
where $\tilde g_\tau$ obtains from  $\tau=[n_\beta]$ by taking a factor $\psi_\alpha$ for every $n_\alpha=1$, and a factor $1$ for $n_\alpha=0$, with an appropriate sign such that $\tilde g_\tau g_\tau=\psi_1\psi_2\dots \psi_B$. More formally, we may define $\tilde g_\tau$ by the property
\be\label{26}
\int {\cal D}\psi\tilde g_\tau g_\rho=\delta_{\tau\rho},
\ee
where the functional integral over all Grassmann variables reads
\be\label{27}
\int {\cal D}\psi=\int d\psi_B\dots \int d\psi_2\int d\psi_1.
\ee
The elements $g$ and $\tilde g$ associated to the probability distribution $\{p_\tau\}$ are normalized according to
\be\label{28}
\int {\cal D}\psi~ \tilde g g=\int {\cal D}\psi\sum_{\tau',\tau} q_\tau\tilde g_\tau q_{\tau'} g_{\tau'}=\sum_\tau p_\tau=1.
\ee

\medskip\noindent
{\bf 2. \quad Grassmann operators}

We can realize a classical observable $A$ as an operator ${\cal A}$ acting on $g$ according to 
\be\label{29}
{\cal A} g_\tau=A_\tau g_\tau~,~
\int {\cal D}\psi\tilde g{\cal A} g=
\sum_\tau p_\tau A_\tau=\kl A\kr.
\ee
In particular, the occupation number operator ${\cal N}_\alpha$ associated to $N_\alpha$ reads
\be\label{30}
{\cal N}_\alpha=\partial_\alpha\psi_\alpha=\frac{\partial}{\partial\psi_\alpha}
\psi_\alpha.
\ee
Different occupation numbers commute, ${\cal N}_\alpha{\cal N}_\beta={\cal N}_\beta{\cal N}_\alpha$. The equal-time correlation functions \eqref{13} can be computed as 
\be\label{31}
\bar C^{(m)}_{\alpha_1,\dots\alpha_m}=\int{\cal D}\psi\tilde g{\cal N}_{\alpha_1}\dots 
{\cal N}_{\alpha_m}g.
\ee

The bit operators find now a simple expression. For example, the annihilation operator for bit one, which is represented by eq. \eqref{21} on the level of the states $\tau$, reads now
\be\label{32}
{\cal B}^{(1)}_-=\psi_1~,~{\cal B}^{(1)}_+=\partial_1=\frac{\partial}{\partial\psi_1}.
\ee
We can use standard annihilation and creation operators for fermions
\ba\label{33}
a_\alpha=\psi_\alpha~,~a^\dagger_\alpha=\partial_\alpha~,~
{\cal N}_\alpha=a^\dagger_\alpha a_\alpha,\nn\\
\{a_\alpha,a_\beta\}=\{a^\dagger_\alpha,a^\dagger_\beta\}=0~,~
\{a^\dagger_\alpha,a_\beta\}=\delta_{\alpha\beta}.
\ea
(We use here the standard notation $a^\dagger$ for the creation operators, even though the notion of hermitean conjugation has not been defined for the Grassmann algebra at this stage.) The operators $a_\alpha,a^\dagger_\alpha$ correspond to the bit operators 
${\cal B}^{(\alpha)}_-,{\cal B}^{(\alpha)}_+$ up to a possible minus sign which reflects the position of $n_\alpha$ in the chain of bits for a given element $g_\tau$.

\medskip\noindent
{\bf 3. \quad Grassmann valued wave function}

The element $g=\sum_\tau q_\tau g_\tau$ plays the role of a Grassmann-valued wave function, similar to a vector which obtains as a sum over basis elements $g_\tau$ with coefficients $q_\tau$. The probabilities $p_\tau$ obtain from $g$ by using a projector ${\cal P}_\tau$ with the property
\ba\label{34}
{\cal P}_\tau g&=&q_\tau g_\tau~,~p_\tau=\int {\cal D}\psi\tilde g{\cal P}_\tau g,\\
{\cal P}^2_\tau&=&{\cal P}_\tau~,~\sum_\tau{\cal P}_\tau=1_{{\cal G}},\nn\\
{\cal P}_\tau{\cal P}_\rho&=&{\cal P}_\tau\delta_{\tau\rho}~,
~{\cal P}_\tau g_\rho=g_\tau\delta_{\tau\rho}.\nn
\ea
Here $1_{{\cal G}}$ is the unit operator in the Grassmann algebra, $1_{{\cal G}} g=g$. If we describe $g$ by the set of coefficients $\{ q_\tau\}$ one has 
${\cal P}_\rho\{q_\tau\}=\{0,0,\dots 0,q_\rho,0\dots 0\}$. 

For an arbitrary Grassmann operator ${\cal F}$ one has
\be\label{35}
f={\cal F}g=\sum_\tau f_\tau g_\tau~,~{\cal F} g_\rho =\sum_\tau F_{\tau\rho}g_\tau.
\ee
For linear operators this yields
\be\label{36}
{\cal F}g={\cal F}\sum_\rho q_\rho g_\rho=\sum_\rho q_\rho {\cal F} g_\rho=
\sum_{\tau,\rho} F_{\tau\rho}q_\rho g_\tau,
\ee
and we infer
\be\label{37}
f_\tau=\sum_\rho F_{\tau\rho}q_\rho.
\ee
If we consider the coefficients $\{q_\rho\}$ as real vectors, the linear Grassmann operators are represented as real matrices $F_{\tau\rho}$. 

In this picture the projector ${\cal P}_\sigma$ is represented by the diagonal matrix 
$({\cal P}_\sigma)_{\tau\rho}$, 
\be\label{38}
{\cal P}_\sigma\to({\cal P}_\sigma)_{\tau\rho}=\delta_{\sigma\tau}\delta_{\sigma\rho}.
\ee
For an arbitrary Grassmann operator one has
\be\label{39}
{\cal P}_\tau{\cal F}{\cal P}_\tau={F}_{\tau\tau}{\cal P}_\tau.
\ee
The explicit form of ${\cal P}_\tau$ can be written as
\be\label{40}
{\cal P}_\tau=g_\tau {\cal D}_\tau\tilde{{\cal D}}_\tau\tilde g_\tau,
\ee
where $g_\tau,\tilde g_\tau$ stand for the multiplication with the corresponding Grassmann elements. The operators ${\cal D}_\tau,\tilde{{\cal D}}_\tau$ involve products of derivatives such that applied on $g_\tau$ and $\tilde g_\tau$ they yield 
\be\label{41}
{\cal D}_\tau g_\tau=1~,~\tilde{{\cal D}}_\tau\tilde g_\tau=1.
\ee
Thus the projector ${\cal P}_\tau$ provides an explicit formal construction for the extraction of the probability distribution $\{p_\tau\}$ from the Grassmann wave function $g$. 

At this stage we have expressed all statistical properties of our system of two-level observables in terms of a Grassmann algebra. We can associate to every normalized element of a Grassmann algebra $(\sum_\tau q^2_\tau=1)$ a set of classical probabilities $p_\tau=q^2_\tau$, and to every ``real diagonal'' Grassmann operator ${\cal A}$ a classical observable $A_\tau$ with expectation value given by eq. \eqref{29}. The time evolution of the probability distribution typically involves Grassmann operators ${\cal F}$ \eqref{35} that are not diagonal. In appendix B we discuss in detail the unitary time evolution of sect. \ref{Evolutionlaw} in the Grassmann formalism.

\medskip\noindent
{\bf 4. \quad Complex Grassmann algebra}

The Grassmann algebra can be extended to a complex Grassmann algebra, where the coefficients $q_\tau$ are replaced by complex numbers $c_\tau$, such that the probability distribution $\{p_\tau\}$ is represented by
\be\label{42}
g=\sum_\tau c_\tau g_\tau~,~\tilde g =\sum_\tau c^*_\tau\tilde g_\tau~,~
p_\tau=|c_\tau|^2.
\ee
As discussed before, we may use the last bit for the distinction between the real and imaginary parts of a complex wave function, with $\alpha=(k,B)$ and $\tau'$ labeling the $2^{B-1}$ states corresponding to the bits labeled by $k$,
\ba\label{51A}
q_{\tau',+}&=&q\Big(\big[n_k\big],n_B=1\Big)~,~q_{\tau',-}=q
\Big(\big[n_k\big],n_B=0\Big),\nn\\
c_{\tau'}&=&q_{\tau',_+}+iq_{\tau',-}.
\ea
We may now construct a complex Grassmann algebra using Grassmann variables $\psi_k$ (no $\psi_B$!) and basis elements $g_{\tau'}$, with 
$p_{\tau'}=|c_{\tau'}|^2$ and $g=c_{\tau'}g_{\tau'},\tilde g =c^*_{\tau'}\tilde g_{\tau'}$. In principle, a Grassmann algebra based on $B-1$ Grassmann variables with complex coefficients $c_{\tau'}$ contains the information about $B$ bits. However, the probabilities $p_{\tau'}$ correspond to a type of ``coarse graining'' since they add the probabilities of states with $n_B=1$ or $n_B=0,p_{\tau'}=q^2_{\tau',+}+q^2_{\tau',-}$. 

Correspondingly, only the occupation numbers ${\cal N}_k$ are represented by eq. \eqref{30} , whereas ${\cal N}_B$ projects on the real part of the Grassmann algebra, 
${\cal N}_B(\sum_\tau c_\tau g_\tau)=\sum_\tau Re(c_\tau)g_\tau$. (Note that at this stage the notion of ``real'' or ``complex'' is not defined for the Grassmann variables $\psi_k$ or basis elements $g_{\tau'}$.) The multiplication of a Grassmann element by the imaginary unit $i$ corresponds to a bit operator acting on $n_B$, namely $q(n_B=1)\to q(n_B=0),q(n_B=0)\to -q(n_B=1)$, or a switch between  $n_B=1$ and $n_B=0$, with a subsequent change of sign for the states $q(n_B=1)$. We emphasize that the association \eqref{51A} defines a particular complex structure. Many other complex structures are possible for the wave function and the Grassmann algebra. 

\section{Fermions}
\label{Fermions}

In this short section we recall how each element of the Grassmann algebra can be associated with the quantum wave function for a multi-fermion system. In turn, this implies that the local probabilities $\{p_\tau(t)\}$ can be used in order to describe fermionic quantum systems. Of course, this is only one of the possible interpretations. The formulation in terms of Grassmann elements is particularly suitable for a system of fermions.

A pure state quantum wave function for a multi-fermion system with $F$ different fermion species can be written in the occupation number basis as 
\be\label{FA1}
\varphi=\sum_\tau c_\tau\hat\varphi_\tau~,~\sum_\tau|c_\tau|^2=1,
\ee
with basis states
\ba\label{FA2}
\hat\varphi_1&=&|1,1,1\dots 1\kr~,~\hat\varphi_2=
|0,1,1,\dots 1\kr,\nn\\
\hat\varphi_3&=&|1,0,1\dots 1\kr,\dots
\ea
Here the index $\tau$ labels the $2^F$ basis states with fixed occupation numbers and $c_\tau$ are complex coefficients. We can associate the basis states $\hat\varphi_\tau$ with the basis elements $g_\tau$ of the Grassmann algebra
\be\label{FA3}
g_1=1~,~g_2=\psi_1~,~g_3=\psi_2,\dots
\ee
For this purpose we work with a complex Grassmann algebra, with $F=B-1$, and the last bit labeling real and imaginary parts of the wave function. The wave function $\varphi$ \eqref{FA1} is then associated with an element of the Grassmann algebra or Grassmann wave function
\be\label{FA4}
\varphi~\hat{=}~g=\sum_\tau c_\tau g_\tau.
\ee

The annihilation and creation operators $a_\alpha,a^\dagger_\alpha$ for fermion species $\alpha$ are represented as Grassmann operators according to eq. \eqref{33}. The basis states can be obtained by applying products of $a_\alpha$ on the totally occupied state $\hat\varphi_1$. (We choose here sign conventions for the basis states such that the sign is positive if the $a_\alpha$ are ordered such that smaller $\alpha$ are to the left. This corresponds to a convention for the $g_\tau$ with positive sign if the factors $\psi_\alpha$ are ordered with the smaller $\alpha$ on the left. We  may call this the ``hole convention'' since the representation of states with a certain number of zeros in the bit chain is particularly simple. In sect. \ref{Quantumfieldtheory} we will encounter a different ``fermion convention''.) In this way all operators acting on Grassmann elements are directly related to operators acting on the multi-fermion wave function.

In particular, we can write the unitary evolution law in the form of a Grassmann evolution operator ${\cal U}(t,t_0)$ acting on an initial state $g(t_0)$
\be\label{FA5}
g(t)={\cal U}(t,t_0)g(t_0).
\ee
Properties of this evolution operator are discussed in the appendix. In the Grassmann formulation the Hamilton operator ${\cal H}$ is defined as
\be\label{113A}
{\cal H}(t)=i\partial_t{\cal U}(t,t_0)\tcu(t,t_0),
\ee
with $\tcu$ the hermitean conjugate of ${\cal U}$ (see appendix B). The time evolution \eqref{FA5} can then be written as a Schr\"odinger equation
\ba\label{113B}
i\partial_tg(t)&=&i\partial_t{\cal U} (t,t_0)g(t_0)\nn\\
&=&{\cal H}(t){\cal U}(t,t_0)g(t_0)={\cal H}(t)g(t).
\ea

\section{Two-state quantum mechanics from unitary evolution of classical probabilities}
\label{Two-state}

So far our discussion has remained rather abstract and general. In the following sections we will discuss a few simple examples. We start in this section by presenting an explicit unitary evolution law for classical probabilities that realizes the Schr\"odinger equation for two-state quantum mechanics.

\medskip\noindent
{\bf 1.\quad Unitary evolution}

We  take a system with only one value of $\alpha$ such that the probability distribution $\big\{p_\tau(t)\big\}$ reduces at a given $t$ to the two probabilities $\big\{p_0(t)~,~p_1(t)\big\}$ for the states with $N(t)=0,1$, with $p_0(t)+p_1(t)=1$. The state of the local-time subsystem is specified by the sequence $\{p_1(t)\}$. We employ here a language with continuous time $t$, but discrete time steps $t_i$ can be easily implemented. We start with the equivalence class of all-time probabilities $\{p_\omega\}$ that lead to the sequence of local-time probabilities $p_1(t)=\cos^2\omega t$ in eq. \eqref{15L}, for which we have constructed some explicit representative all-time probability distributions in sect. \ref{timesubsystem}. The local-time probability distribution obeys the evolution law \eqref{36A}, which belongs to the unitary evolution laws of the type \eqref{21A}, \eqref{21C}. Other sequences of local-time probabilities that obey the same evolution law are given by 
\ba\label{61}
p_1(t)=\cos^2(\omega t)p_1(0)+\sin^2(\omega t)p_0(0)\nn\\
-2s\cos (\omega t)\sin(\omega t)
\sqrt{p_1(0)p_0(0)},
\ea
where we set $t_0=0$ and $s=\pm 1$. For $p_0(0)=0$ this corresponds to the sequence of local-time probabilities \eqref{15L}, while the other states of the local-time subsystem with $p_0(0)\neq 0$ can be constructed from suitable all-time probability distributions $\{p_\omega\}$ in an analogous way. We observe that a change in the sign $s$ is equivalent to a time reflection $t\to -t$.  The mean occupation number obeys
\be\label{61A}
\kl N(t)\kr=p_1(t).
\ee

For an explicit specification of the evolution law the coefficients $V_{\tau\rho\sigma}(t,0)$ in eq. \eqref{21A} read
\ba\label{62}
V_{111}&=&V_{000}=\cos^2(\omega t)~,~V_{100}=V_{011}=\sin^2(\omega t),\\
V_{110}&=&V_{101}=-V_{010}=-V_{001}=-s\cos(\omega t)\sin(\omega t).\nn
\ea
In turn, the rotation matrix $R$ in eq. \eqref{21C} can be written as a real unitary matrix $U$
\be\label{63}
R=\left(\begin{array}{ll}
\cos \omega t,&-\sin \omega t\\\sin\omega t,&\cos \omega t
\end{array}\right)=U=\exp (-i\omega\tau_2 t).
\ee
This corresponds to two-state quantum mechanics with a Hamiltonian $H=\omega\tau_2$ and a real two-component initial wave function $\varphi(0)=\big(q_1(0),q_0(0)\big)=(\sqrt{p_1(0)},s\sqrt{p_0(0)})$. 
Using the standard quantum formalism for a real two-component wave function one has
\ba\label{117A}
\varphi(t)=\left(\begin{array}{l}{q_1(t)}\\{q_0(t)}\end{array}\right)~,~\partial_t\varphi(t)=-iH\varphi(t)~,
~H=\omega\tau_2.
\ea
The wave function $\varphi(t)=U(t,0)\varphi(0)$ remains real in the course of the evolution. Evaluating the expectation value of the operator $\hat N=\frac12(1+\tau_3)$, which is associated to the occupation number observable,
\ba\label{64}
\kl N(t)\kr=\kl\varphi(t)|\hat N|\varphi(t)\kr=
\kl\varphi(0)U^\dagger(t,0)|\hat N|U(t,0)\varphi(0)\kr,\nn\\
\ea
one recovers eq. \eqref{61A}.

\medskip\noindent
{\bf 2. \quad Evolution of Grassmann elements}

The Grassmann algebra (at a given $t$) contains only two basis elements, $\{g_\tau\}=\{g_1,g_0\}=\{1,\psi\}$ and the state is represented by
\be\label{88A}
g(t)=q_1(t)+q_0(t)\psi,
\ee
with

\ba\label{88B}
q_1(t)&=&\cos(\omega t)\sqrt{p_1(0)}-\sin(\omega t)s\sqrt{p_0(0)},\nn\\
q_0(t)&=&\sin (\omega t)\sqrt{p_1(0)}+\cos(\omega t)s\sqrt{p_0(0)},
\ea
and
\be\label{65}
p_1(t)=q_1(t)^2~,~p_0(t)=q_0(t)^2.
\ee
We observe that a given initial sign convention for $q_1(0)$ and $q_0(0)$ fixes the signs for $q_1(t)$ and $q_0(t)$ for all $t$. The sign convention at $t=0$ is arbitrary and corresponds to a choice of gauge. 

The Grassmann evolution operator
\ba\label{66}
{\cal U}(t)&=&\cos(\omega t)+\sin (\omega t)\left(\psi-\frac{\partial}{\partial \psi}\right)\nn\\
&=&\cos(\omega t)+\sin (\omega t)(a-a^\dagger)
\ea
describes the time evolution
\be\label{67}
g(t)={\cal U}(t)g(0).
\ee
The conjugate basis elements read $\{\tilde g_\tau\}=\{\psi,1\}$, with
\ba\label{68}
\tilde g(t)&=&q_0(t)+q_1(t)\psi.
\ea
Correspondingly, the conjugate evolution operators read
\ba\label{69}
\tilde {\cal U}^T(t)&=&\cos(\omega t)-\sin 
(\omega t)\left(\psi+\frac{\partial}{\partial\psi}\right){\cal S}\nn\\
\tcu(t)&=&\cos(\omega t)-\sin(\omega t)\left(\psi-\frac{\partial}{\partial\psi}\right),\nn\\
\tilde {\cal U}(t){\cal U}(t)&=&1.
\ea
Here we use the sign operator ${\cal S}$ which reverses the sign of all odd elements of the Grassmann algebra, i.e.
\be\label{69A}
{\cal S}g_\tau=(-1)^{M_\tau}g_\tau,
\ee
with $M_\tau$ the number of factors of $\psi$ in $g_\tau$. 

Using $\left(\psi-\frac{\partial}{\partial\psi}\right)^2=-1$ we can write
\ba\label{70}
{\cal U}(t)&=&\exp\left\{\omega
\left(\psi-\frac{\partial}{\partial\psi}\right)t\right\},\nn\\
\tcu(t)&=&\exp\left\{-\omega\left(\psi-\frac{\partial}{\partial\psi}\right)t\right\}.
\ea
This can be cast into a Hamiltonian form similar to quantum mechanics,
\ba\label{71}
i\partial_t{\cal U}&=&{\cal H}{\cal U}~,~{\cal U}=\exp (-i{\cal H}t),\nn\\
{\cal H}&=&i\omega
\left(\psi-\frac{\partial}{\partial\psi}\right)=i\omega(a-a^\dagger),
\ea
with
\ba\label{72}
a=\left(\begin{array}{l}0,0\\1,0\end{array}\right)~,~
a^\dagger=\left(\begin{array}{l}0,1\\0,0\end{array}\right).
\ea
This realizes the Schr\"odinger equation $i\partial_tg={\cal H}g$. 

\medskip\noindent
{\bf 3. \quad Time evolution of occupation number}

In the Grassmann formulation the occupation number is represented by the Grassmann operator
\be\label{73}
{\cal N}=\frac{\partial}{\partial\psi}\psi=a^\dagger a.
\ee
It translates to the quantum operator $\hat N=(1+\tau_3)/2$, with
\ba\label{74}
\kl N(t)\kr &=&\int d\psi\tilde g(t)\frac{\partial}{\partial\psi}\psi g(t)\\
&=&\kl\varphi(t)|\hat N|\varphi(t)\kr
=\varphi^\dagger(t)\frac{1+\tau_3}{2}\varphi(t).\nn
\ea
Motivated by the quantum analogue we evaluate further Grassmann operators as ${\cal A}=\psi+\frac{\partial}{\partial \psi}$,
\ba\label{75}
\kl A(t)\kr&=&\int d\psi\tilde g(t)\left(\psi+\frac{\partial}{\partial\psi}\right)
g(t)=
2q_0(t)q_1(t)\nn\\
&=&2\sin(\omega t)\cos(\omega t)(2p_{1,0}-1)\\
&&+2s\big (\cos^2(\omega t)-\sin^2(\omega t)\big)\sqrt{p_{1.0}(1-p_{1,0})}.\nn
\ea

A priori, the classical statistical interpretation of $\kl A(t)\kr$ may not seem obvious. A quick calculation reveals, however, that $\kl A\kr$ coincides with the time derivative of eq. \eqref{61}  up to a factor $-\omega$,
\be\label{76}
\partial_t\kl N\kr=-\omega\kl A\kr.
\ee
This is not surprising in the quantum picture. The quantum operator associated to ${\cal A}$ is $\tau_1$, and we may compute $\partial_t\kl N\kr$ in the Heisenberg picture, $[H,\hat N]=i\omega\tau_1$, or
\ba\label{77}
\partial_t\kl N\kr=i\kl\varphi(t)[H,\hat N]\varphi(t)\kr
=-\omega\kl \varphi(t)|\tau_1|\varphi(t)\kr.
\ea

This simple observation has interesting implications. We have started with a simple classical statistical ensemble, say with a probability distribution \eqref{15H}, \eqref{15K} or the generalization to a finer grid discussed in sect. \ref{all-time}. The local (in time) probabilities $p_0(t),p_1(t)$ show the periodic behavior \eqref{15L}. This classical statistical ensemble is characterized by the sequences of local-time probabilities \eqref{61} which obey a simple unitary evolution law. Despite this completely classical statistical structure, we have found a quantum description with non-commuting observables
\be\label{78}
\hat N=\frac{1}{2}(1+\tau_3)~,~\partial_t\hat N=-\omega\tau_1.
\ee
We will discuss this issue in the next section.

\section{Unequal time correlation functions and non-commuting quantum operators}
\label{Unequaltimecorrelation}

We next address the question which properties follow purely from the evolution law of the classical probabilities, and which ones need additional input from the all-time probability distribution. This will concern unequal time correlations of the type \eqref{3} and the status of the observables as $\partial_tN(t)$. 

\medskip\noindent
{\bf 1. \quad Time-derivative observables}

The status of the observables $N$ and $\partial_t N$ is different. Going back to the formulation with discrete time steps it is obvious that $\kl N(t_i)\kr$ can be directly computed from the probability $p_1(t_i)$, where the past and future play no role and can be integrated out according to eq. \eqref{12}. In contrast, the expectation value of the observable
\be\label{79}
(\partial_t N)_\epsilon(t_i)=\frac{N(t_i)-N(t_i-\epsilon)}{\epsilon}
\ee
requires, in principle, more information than contained in $p_1(t_i)$. It involves the joint probabilities of two~``time layers''~$t_i$~and~$t_i-\epsilon$, corresponding to extended local probabilities for an interval with two time points in the sense of sect. \ref{timesubsystem}. One needs the four probabilities $\big\{p_{11}(t_i,t_i-\epsilon)$, $p_{10}(t_i,t_i-\epsilon),$ $p_{01}(t_i,t_i-\epsilon)$, $p_{00}(t_i,t_i-\epsilon)\big\}$ for finding $N(t_i)$ and $N(t_i-\epsilon)$ in the combinations $(1,1), (1,0), (0,1)$ and $(0,0)$, respectively. (The sum of the four joint probabilities equals one.) They can be obtained from eq. \eqref{15A}, now integrating out only the future $(t'>t_i)$ and the past before $t_i-\epsilon~(t'<t_i-\epsilon)$. 

In terms of the joint probabilities one finds $(\partial_tN)_\epsilon(t_i)$ and the classical correlation between $N(t_i)$ and $(\partial_tN)_\epsilon(t_i)$ as 
\ba\label{80}
\kl(\partial_tN)_\epsilon(t_i)\kr&=&\frac1\epsilon(p_{10}-p_{01}),\\
\kl N(t_i)(\partial_tN)_\epsilon(t_i)\kr&=&\frac1\epsilon p_{10}.\label{81}
\ea
Two out of three independent joint probabilities can be computed from $p_1(t_1)$ and $p_1(t_i-\epsilon)$,
\ba\label{82}
\kl N(t_i)\kr&=&p_{11}+p_{10}=p_1(t_i),\nn\\
\kl N(t_i-\epsilon)\kr&=&p_{11}+p_{01}=p_1(t_i-\epsilon).
\ea
This allows us to compute $p_{10}-p_{01}$ and therefore $\kl(\partial_tN)_\epsilon(t_i)\kr$ from $p_1(t_i)$ and $p_1(t_i-\epsilon)$, in contrast to $p_{10}$ and the correlation $\kl N(\partial_t N)_\epsilon\kr$. The expectation value of $(\partial_tN)_\epsilon$ can therefore be computed from the information contained in the state of the local-time subsystem, while $\kl N(\partial_tN)_\epsilon\kr$ needs information beyond the  local-time subsystem. 

The evolution law relates $p_1(t_1-\epsilon)$ to $p_1(t_1)$
\ba\label{83}
&&\hspace{-0.5cm}p_1(t_i)=\sum_{\rho,\sigma=0,1}
V_{1\rho\sigma}(t_i,t_i-\epsilon)
\sqrt{p_\rho(t_i-\epsilon)p_\sigma(t_i-\epsilon)}\nn\\
&&=\sum_{\rho,\sigma}R_{1\rho}(t_i,t_i-\epsilon)R_{1\sigma}
(t_i,t_i-\epsilon)q_\rho(t_i-\epsilon)q_\sigma(t_i-\epsilon)\nn\\
&&=\cos^2(\omega\epsilon)p_1(t_i-\epsilon)+\sin^2
(\omega\epsilon)p_0(t_i-\epsilon)\nn\\
&&\quad -2\cos(\omega\epsilon)\sin(\omega\epsilon)q_1(t_1-\epsilon)q_0(t_i-\epsilon),
\ea
and we infer
\ba\label{84}
&&\hspace{-0.5cm}\kl (\partial_tN)_\epsilon(t_i)\kr=-\frac2\epsilon\cos
(\omega\epsilon)\sin(\omega\epsilon)\nn\\
&&\sqrt{p_1(t_1-\epsilon)\big (1-p_1(t_i-\epsilon)\big)}
\text{sign}\big((q_1q_0)(t_i-\epsilon)\big)\nn\\
&&+\frac1\epsilon\sin^2(\omega\epsilon)\big (1-2p_1(t_i-\epsilon)\big).
\ea
In the limit $\epsilon\to 0$ only the first term contributes, yielding
\ba\label{85}
\kl\partial_t N(t)\kr&=&-2\omega\sqrt{p_1(t)\big (1-p_1(t)\big)}s(t),\nn\\
s(t)&=&\text{sign}\big (q_1(t)q_0(t)\big).
\ea
By virtue of the evolution law $\kl \partial_tN\kr$ is almost computable in terms of the probability $p_1(t)$ - the only additional information which keeps some additional memory of the states of the system at earlier times is its sign $s(t)$. The wave function $(q_1,q_0)$ precisely contains the sign information, beyond the probabilities related to $|q_1|,|q_0|$. (An overall common sign of $q_0$ and $q_1$ is unobservable since it cancels out in the computation of observables.)

We may interpret the pair $\big(p_1(t),s(t)\big)$ as the information characterizing the state of a statistical ``fixed-time subsystem'' related to a given time $t$. Information contained in the full ensemble for all $t_i$ - i.e. in the sequence of all local-time probabilities $\big\{p_1(t_i)\big\}$ - which goes beyond $\big(p_1(t),s(t)\big)$ is not relevant for expectation values of fixed-time system observables. The information $\big(p_1(t),s(t)\big)$ which characterizes the state of the fixed-time subsystem is conveniently collected in the wave function $q_\tau(t),\tau=0,1$. Apart from the irrelevant overall sign the information about the wave function is indeed contained in  $p_1(t)$ and $s(t)$. Similar to quantum mechanics it is nevertheless convenient to specify the fixed-time subsystem with the two-component normalized wave function $q_\tau(t)$. This contains the overall sign as redundant information, similarly to the redundant overall phase in quantum mechanics. For any initial condition $q_\tau(t_0)$ the fixed-time subsystem is completely determined by the evolution law. In terms of the wave function eq. \eqref{85} finds the simple expression 
\be\label{119A}
\kl \partial_t N(t)\kr =-2\omega q_1(t) q_0(t).
\ee

\medskip\noindent
{\bf 2. \quad Ambiguities for time-derivative observables}

Even though $\kl\partial_t N\kr$ can be computed from the wave function $\{q_\tau(t)\}$ it is not a system observable in the standard sense. The expectation value of the squared classical observable 
$\big (\partial_tN(t)\big)^2$ cannot be computed unambiguously from $\big(p_1(t),s(t)\big)$. This becomes apparent if we go back to discrete time steps and consider $(\partial_tN)_\epsilon$. The spectrum of the possible values of $(\partial_tN)_\epsilon$ in a classical state $\omega=\{n(t_i)\}$ contains three values $(-\frac1\epsilon,0,\frac1\epsilon)$. Thus $(\partial_tN)^2_\epsilon$ has eigenvalues $1/\epsilon^2$ and $0$ and the expectation value
\be\label{168}
\kl (\partial_tN)^2_\epsilon\kr=\frac{1}{\epsilon^2}(p_{10}+p_{01})
\ee
diverges for $\epsilon\to 0$ (unless $p_{10}+p_{01}\sim \epsilon^2$, which is only possible for $\kl \partial_tN\kr=0$).

Furthermore, there are several different discrete observables which yield the same continuum limit for $\partial_t N$. For example, $\partial_tN$ can be associated with $(\partial_tN)^-_\epsilon$ as given by eq. \eqref{79}, but also with 
\be\label{113AA}
(\partial_tN)_\epsilon^+=\frac{N(t_i+\epsilon)-N(t_i)}{\epsilon}
\ee
or
\ba\label{113Aa}
(\partial_tN)_{2\epsilon}&=&\frac{N(t_i+\epsilon)-N(t_i-\epsilon)}{2\epsilon}\nn\\
&=&\frac12\big((\partial_tN)^+_\epsilon+(\partial_tN)^-_\epsilon\big).
\ea
The expectation value of $(\partial_tN)^+_\epsilon$ obeys the same formula as for $(\partial_tN)^-_\epsilon$, except that the wave function is now taken at $t+\epsilon$ instead of $t$. This difference vanishes in the continuum limit such that the expectation values of the discrete observables \eqref{79}, \eqref{113AA} and \eqref{113Aa} have a common continuum limit. However, we can now also consider products as
\ba\label{113Ba}
(\partial_tN)^+_\epsilon(\partial_tN)^-_\epsilon&=&\frac{1}{\epsilon^2}\big\{N(t_i)N(t_i+\epsilon)+N(t_i)N(t_i-\epsilon)\nn\\
&&-N(t_i+\epsilon)N(t_i-\epsilon)-N(t_i)\}.
\ea
The expectation value of this observables involves joint probabilities at three time layers $t_i+\epsilon,t_i$ and $t_i-\epsilon$. Using the relations
\ba\label{113Ca}
\kl N(t_i)\kr&=&p_{111}+p_{110}+p_{011}+p_{010},\nn\\
\kl N(t _i)N(t_i+\epsilon)\kr&=&p_{111}+p_{110},\nn\\
\kl N(t_i)N(t_i-\epsilon)\kr&=&p_{111}+p_{011},\nn\\
\kl N(t_i+\epsilon)N(t_i-\epsilon)\kr&=&p_{111}+p_{101},
\ea
one obtains a negative expectation value 
\be\label{113D}
\kl(\partial_tN)^+_\epsilon (\partial_tN)^-_\epsilon\kr=-\frac{1}{\epsilon^2}(p_{101}+p_{010}),
\ee
in contrast to the positive expectation values
\ba\label{113E}
\kl \big[(\partial_t N)^+_\epsilon\big]^2\kr&=&\frac{1}{\epsilon^2}(p_{101}+p_{100}+p_{011}+p_{010})\nn\\
\kl \big[(\partial_t N)^-_\epsilon\big]^2\kr&=&\frac{1}{\epsilon^2}(p_{110}+p_{010}+p_{101}+p_{001}).
\ea
The discrete observables \eqref{113D} and \eqref{113E} take different values for $\epsilon\to 0$ and there is no unambiguous well defined continuum limit for $\kl(\partial_tN)^2\kr$. 

While the continuum limit of the expectation value of the observable $\partial_t N$ is well defined, the spectrum of the discrete observables $(\partial_tN)^\pm_\epsilon$ and $(\partial_tN)_{2\epsilon}$ is different - for the last observable the possible values for the classical states are $(\partial_tN)_{2\epsilon}=\pm 1/2\epsilon,0$. This lack of uniqueness of the spectrum is in close correspondence with the lack of a well defined continuum limit for $(\partial_tN)^2$. If one would like to perform measurements of $\partial_tN$ or $(\partial_tN)^2$ the lack of uniqueness of the discrete observables would be very annoying. The outcome of a series of measurements in identical settings would depend crucially on the precise details of the measurement apparatus which would need a time resolution of the order $\epsilon$ and a precise statement which particular discrete observable is measured by a particular apparatus. These are typically not properties of a realistic apparatus, and one would like to have a more robust definition of observables that can be measured. That this should be possible is suggested by the existence of a unique limit for $\kl \partial_t N\kr$. 

\medskip\noindent
{\bf 3. \quad Non-commuting quantum operators}

The lack of robustness of observables containing time derivatives has already been noted in ref. \cite{3}. There it was proposed that more robust observables can be based on equivalence classes that correspond to the operators in quantum mechanics. Indeed, we have found in eq. \eqref{78} the relation
\be\label{169}
\kl \partial_tN\kr=-\omega\kl \tau_1\kr=-2\omega q_1(t)q_0(t).
\ee
This equals precisely the expression \eqref{85}. (We can identify in leading order $t-\epsilon$ and $t$.) The information contained in the wave function $q(t)$ is sufficient to compute the expectation values of off-diagonal operators as $\tau_1$ in addition to the diagonal ones. 

Observables as $\partial_tN$ can be represented as commuting ``diagonal observables'' for the all-time statistical  ensemble for all states $\big\{n(t)\big\}$. While the spectrum depends on the details of the particular definition, the expectation value is robust in the sense that the detailed implementation does not matter. The robust expectation value can also be computed from off-diagonal quantum operators for the subsystem at a given time. In accordance with the general considerations how a quantum  system can arise from a classical statistical ensemble the joint probabilities for two observables expressed by non-commuting operators are not available within the subsystem \cite{QPSS}. In our case this concerns the classical correlation $\kl N\cdot \partial_t N\kr$ which cannot be computed from the information in the local-time subsystem. 

The robustness of the expectation value raises the interesting question if one can find a classical observable for $\partial_t N$ that becomes a system observable in the continuum limit. Such an observable must be a two-level observable with possible measurement values $\omega$ and $-\omega$. We leave this for future investigation.

\medskip\noindent
{\bf 4. \quad Unequal time occupation numbers}

While the issue of a useful classical observable for $\partial_t N$ is somewhat involved, there are other much simpler circumstances where off-diagonal quantum operators appear for the local subsystem at a given time $t$. The classical observable $N(t+\alpha)$ is well defined, with spectrum $(1,0)$. For $\alpha\neq 0$ it may be called somewhat loosely an ``unequal time occupation number'', where unequal relates to different time arguments for the occupation number and the local-time probabilities. The expectation value $\kl N(t+\alpha)\kr$ is computable from the information available for the local subsystem at time $t$, characterized by $q_\tau(t)$ or $\varphi(t)$. One has the identity
\ba\label{114A}
\kl N(t+\alpha)\kr&=&q^2_1(t+\alpha)\\
&=&\frac12\varphi^T(t)
R^T(\alpha) (1+\tau_3)R(\alpha)\varphi(t),\nn
\ea
with, cf. eq. \eqref{63},
\be\label{114B}
R(\alpha)=\left(\begin{array}{ccc}\cos \omega\alpha&,&-\sin\omega\alpha\\\sin\omega\alpha&,&\cos\omega\alpha\end{array}\right)=
U(t+\alpha,t).
\ee
We can therefore express the expectation value of $N(t+\alpha)$ by the quantum operator $\hat N(t+\alpha)$,
\ba\label{114C}
\kl N(t+\alpha)\kr &=&\varphi^T(t)\hat N(t+\alpha)\varphi(t),\nn\\
\hat N(t+\alpha)&=&\frac12 R^T(\alpha)(1+\tau_3)R(\alpha)\\
&=&\frac12 U^\dagger(t+\alpha,t)(1+\tau_3)U(t+\alpha,t).\nn
\ea
This is precisely the Heisenberg picture of quantum mechanics, cf. eq. \eqref{64}. Since $N(t+\alpha)$ is a classical two-level observable and its expectation value is computable from the state of the local-time subsystem it is a system observable. 

The interpretation of measurements in quantum mechanics can now be inferred from the basic postulates of classical statistical mechanics. In the all-time classical statistical ensemble $N(t+\alpha)$ is a standard classical observable, and the only possible outcomes of measurements are zero and one. This corresponds to the spectrum of eigenvalues of the quantum operator $\hat N(t+\alpha)$, explaining the quantum rule that only eigenvalues of the quantum operator $\hat A$ associated to an observable will be found in measurements. The quantum rule for the expectation value \eqref{114C} has been obtained from the classical rule for expectation values. The probabilistic interpretation of quantum mechanics states that an eigenvalue $\lambda_m$ will be found with a probability $p_m$. This probability obtains by choosing a basis $|m\kr$ of eigenvalues of $\hat A$, $\hat A|m\kr=\lambda_m|m\kr$, as 
\be\label{114D}
p_m=|\kl \varphi|m\kr|^2.
\ee
(We assume here non-degenerate eigenvalues for simplicity.) For the operator $\hat N(t+\alpha)$ these are exactly the probabilities in the classical statistical ensemble to find the values $\lambda_1=1,\lambda_2=0$ of the observable $N(t+\alpha)$, namely
\ba\label{114E}
w_1&=&p_1(t+\alpha)=q^2_1(t+\alpha)\nn\\
&=&\big(\varphi_\tau(t)R_{1\tau}(\alpha)\big)^2=|\kl\varphi(t)|1\kr|^2,
\ea
with 
\be\label{114F}
|1\kr=\left(\begin{array}{l}R_{11}(\alpha)\\R_{12}(\alpha)\end{array}\right)~,~|2\kr=
\left(\begin{array}{l}R_{21}(\alpha)\\R_{22}(\alpha)\end{array}\right),
\ee
and $w_2=1-w_1$. We conclude that non-commuting operators and their standard association to observables and measurements in quantum mechanics arise in a natural way from our classical statistical ensemble if one focuses on local-time subsystems.

The explicit operator representing the occupation number $N(t+\alpha)$ in the local-time system with state characterized by $\varphi(t)$ reads
\ba\label{114FX}
&&\hat N(t+\alpha)=\left(\begin{array}{ccc}
\cos^2\omega\alpha&,&-\sin\omega\alpha\cos\omega\alpha\\
-\sin\omega\alpha\cos\omega\alpha&,&\sin^2\omega\alpha 
\end{array}\right)\nn\\
&&=\frac12+\frac12(\cos^2\omega\alpha-\sin^2\omega\alpha)\tau_3-\sin\omega\alpha\cos\omega\alpha\tau_1.\nn\\
\ea
In particular, one finds
\be\label{114FY}
\hat N\left(t+\frac{\pi}{4\omega}\right)=\frac12(1-\tau_1).
\ee
This yields for the expectation value of $\partial_tN$ the relation
\be\label{114FZ}
\kl \partial_t N(t)\kr=2\omega\left (\kl \hat N\left(t+\frac{\pi}{4\omega}\right)\kr-\frac12\right ).
\ee
This relation holds independently of the particular definition of the observable $\partial_tN$, while the observables $\partial_tN(t)$ and $N\left(t+\frac{\pi}{4\omega}\right)$ are not necessarily identical. 

\section{Measurements and quantum correlation}
\label{Measurements and quantum correlation}

The classical correlation function $\kl N(t+\alpha)\cdot N(t)\kr$ cannot be computed from the information of the local-time subsystem. This conforms with the general discussion of ref. \cite{QPSS} and reflects the fact that the local-time subsystem is described by incomplete statistics. We can therefore not use the classical correlation function or joint probabilities for the description of a sequence of idealized measurements in the subsystem. Such sequences of measurements have to be described by a different ``measurement correlation'' which is based on conditional probabilities rather than joint probabilities \cite{QPSS}. The measurement correlation $\kl N(t+\alpha)N(t)\kr_m$ is given by the conditional probability to find $N(t+\alpha)=1$ {\em if} $N(t)$ has been found to be equal to one. For an idealized measurement in a subsystem the conditional probability must be computable in terms of the available information $\varphi(t)$. It should not involve additional information about the ``environment'', as the classical correlation does. 

For a two-state system a consistent conditional probability obeying this criterion leads to 
\be\label{114G}
\kl N(t+\alpha)N(t)\kr_m=
\big(\hat N(t+\alpha)\big)_{11}\big(q_1(t)\big)^2.
\ee
This can be expressed by the rule that after the first measurement $N(t)=1$ the wave function $\varphi(t)$ is replaced by the eigenstate of $\hat N(t)$ with eigenvalue one, and that subsequently $N(t+\alpha)$ is measured in this new state. This is the ``reduction of the wave function'', which appears here merely as a convenient rule for specifying the conditional probability relevant for a sequence of idealized measurements in a subsystem. 

The rule of ``reduction of the wave function'' for the computation of conditional probabilities can be generalized to other system observables. It  is consistent with the requirement that two measurements of an observable $A$ at the same time $t$ (or at infinitesimally close times) should yield the same measurement value. We will assume in the following observables with a non-degenerate spectrum. If the first measurement yields $\lambda_m$, the ``reduced state'' is $|m\kr$. Then the expectation value of $A$ in the reduced state is again $\lambda_m$, and
\ba\label{114I}
\kl A(t)A(t)\kr_m&=&\lambda^2_m|\kl \varphi(t)|m\kr|^2\nn\\
&=&\kl\varphi(t)\hat A^2(t)\varphi(t)\kr.
\ea

More generally, a consistent conditional probability to find the value $\sigma_{n_B}$ of an observable $B$ if $\lambda_{m_A}$ has been found for $A$ reads
\be\label{114J}
w(\sigma_{n_B}|\lambda_{m_A})=|\kl n_B|m_A\kr|^2,
\ee
where $\hat A|m_A\kr=\lambda_{m_A}|m_A\kr~,~\hat B|n_B\kr=\sigma_{n_B}|n_B\kr$. With $w_{m_A}=|\kl \varphi|m_A\kr|^2$ the probability to find $\lambda_{m_A}$ in the first place, the measurement correlation becomes
\ba\label{114K}
\kl BA\kr_m&=&\sum_{m_A,n_B}\sigma_{n_B}\lambda_{m_A}w(\sigma_{n_B}|\lambda_{m_A})w_{m_A}\nn\\
&=&\sum_{m_A}\lambda_{m_A}w_{m_A}\kl m_A|\hat B|m_A\kr.
\ea
The last expression multiplies the value found in the first measurement of $A$ with the probability to find it, and then with the expectation value of $B$ in the state $m_A$, and finally sums over all possible measurement values of $A$. We can write the measurement correlation in the form 
\be\label{114L}
\kl BA\kr_m=\kl \varphi|\hat C^{(2)}_m(\hat B, \hat A)|\varphi\kr,
\ee
with $\hat C^{(2)}_m$ a hermitean operator (symmetric $2\times 2$ matrix) constructed from the operators $\hat A$ and $\hat B$. In general, the sequence of two measurements matters
\be\label{114M}
\hat C^{(2)}_m(\hat A,\hat B)\neq \hat C^{(2)}_m(\hat B,\hat A).
\ee

For the special case where $\hat A$ and $\hat B$ are traceless two-level observables, with two non-degenerate eigenvalues $(a,-a)$ or $(b,-b)$, one can express \cite{QPSS} the measurement correlation by the anticommutator of the operators 
\be\label{114N}
\hat C^{(2)}_m(\hat B,\hat A)=\hat C^{(2)}_m(\hat A,\hat B)=\frac12\{\hat A,\hat B\}.
\ee
The anticommutator $\{\hat A,\hat B\}$ involves the operator product. In this case the quantum operator product enters directly in the determination of measurement correlations. This operator product reflects a product structure for observables \cite{QPSS} that is consistent with the structure of equivalence classes of observables. Here we recall that two two-level observables $A_1$ and $A_2$ belong to the same equivalence class if they are both system observables such that their expectation values can be computed from the wave function of the local time system, and if $\kl A_1\kr=\kl A_2\kr$ for all possible wave functions $\varphi(t)$. In this case $A_1$ and $A_2$ are represented by the same quantum operator $\hat A$. (An example for our simple two-state system are $\partial_tN(t)/\omega$ and $2N\left(t+\frac{\pi}{4\omega}\right)-1$, which are both represented by the operator $-\tau_1$.)

For more general observables with a non-degenerate spectrum eq. \eqref{114N} does not hold and $\hat C^{(2)}_m$ becomes a non-linear expression of $\hat A$ and $\hat B$. (One can show that no alternative consistent definition of conditional probabilities is possible which would lead to $\kl N(t+\alpha)N(t)\kr_m=\kl\varphi\big\{\hat N(t+\alpha),\hat N(t)\big\} \varphi\kr/2.$) In a basis of eigenstates of $\hat A$ the operator $\hat C^{(2)}_m(\hat B,\hat A)$ is diagonal
\ba\label{114O}
\hat C^{(2)}_m(\hat B,\hat A)&=&\sum_{m_A}|m_A\kr c_{BA}(m_A)\kl m_A|,\nn\\
c_{BA}(m_A)&=&\sum_{n_B}|\kl n_B|m_A\kr|^2\sigma_{n_B}\lambda_{m_A}.
\ea
The diagonal elements $c_{BA}$ can be computed from $\hat A$ and $\hat B$ independently of the particular representation of these operators. On the other hand, the inverted sequence typically also involves off-diagonal elements in this basis,
\ba\label{144A}
\hat C^{(2)}_m(\hat A, \hat B)&=&\sum_{m_A,m_C}
|m_A \kr c_{AB}(m_A,m_C)\kl m_C|,\nn\\
c_{AB}(m_A,m_C)&=&\\
&&\hspace{-2.0cm}\sum_{n_B,m'_A}\kl m_A|n_B\kr 
\kl n_B|m_C\kr
\kl n_B|m'_A\kr|^2\sigma_{n_B}\lambda_{m'_A}.\nn
\ea

One may use projection operators on eigenvectors of $\hat A$ or $\hat B$
\ba\label{114Ba}
P_{m_A}&=&|m_A\kr \kl m_A|~,~\tilde P_{n_B}=|n_B\kr\kl n_B|,\nn\\
\hat A&=&\sum_{m_A}m_A P_{m_A}~,~\hat B=\sum_{n_B}\sigma_{n_B}\tilde P_{n_B},
\ea
and write
\ba\label{114Ca}
\kl BA\kr_m&=&\sum_{m_A,n_B}\lambda_{m_A}\sigma_{n_B}P_{m_A}\tilde P_{n_B}P_{m_A},\nn\\
\kl AB\kr_m&=&\sum_{m_A,n_B}\lambda_{m_A}\sigma_{n_B}\tilde P_{n_B}P_{m_A}\tilde P_{n_B}.
\ea
Since all quantities $\sigma_{n_B},\lambda_{m_A},\kl m_A|n_B\kr$ and $w_{m_A}$ that appear in eq. \eqref{114K} can be extracted from the quantum operators $\hat A$ and $\hat B$ we can construct $\kl BA\kr_m$ in terms of these operators. This measurement correlation therefore only involves the equivalence classes of observables as it should be. For the special case of two-level observables $\lambda_{m_A}=\pm \lambda,\sigma_{n_B}=\pm \sigma$ one recovers eq. \eqref{114N}, while eq. \eqref{114G} can be expressed as
\ba\label{144D}
\kl N(t+\alpha)N(t)\kr_m=\kl\hat N(t)\hat N(t+\alpha)\hat N(t)\kr.
\ea
The expressions \eqref{114Ca} do not depend on the choice of basis for the representation of the operators.

The use of measurement correlations that differ from the classical correlations is mandatory for idealized measurements of a system that is characterized by incomplete statistics, as the local-time subsystem in our case. This is also the way how ``no-go theorems'' based on Bell's inequalities \cite{Be} for a representation of quantum mechanics by a classical statistical ensemble are circumvented. Indeed, a measurement correlation based on joint probabilities instead of conditional probabilities would imply Bell's inequalities \cite{BS}. The experimental verification of a violation of Bell's inequalities can therefore be used as evidence that only the information for the local-time subsystem is available or used for ideal measurements, rather than the all-time probability distribution. We emphasize that correlation functions that differ from the classical correlation function do not only appear in quantum mechanics. They are genuinely needed for a description of measurements in classical statistical systems as well \cite{QPSS}. It is remarkable that already our simplest example of a periodic all-time probability distribution  \eqref{15L} shows almost all characteristic features of quantum mechanics.

\section{Quantum wave function from classical statistics}
\label{Quantumwave}

We have seen several features of quantum mechanics emerging from our attempt to formulate classical statistical ensembles that admit an evolution law that can accommodate for a periodic time evolution. It may be useful to summarize what we have achieved so far:
\begin{enumerate}
\item We associate to the unitary evolution law for a classical probability distribution a quantum mechanical Schr\"odinger equation.
\item For Ising-type classical statistical ensembles with a unitary time evolution we have formulated a representation in terms of Grassmann variables. This allows a simple representation of fermions and bit operations.
\item We have found that system observables for the local-time subsystem can be represented by quantum operators. These operators show a non-commutative product structure. We have advocated that the correlation of idealized measurements within the local-time subsystem is based on the non-commutative operator product.
\end{enumerate}
At this state it may be useful to ask about the status of the quantum mechanical wave function within our classical statistical setting. 

\medskip\noindent
{\bf 1.\quad Maps between wave function and probabilities}

The association between a Schr\"odinger equation and the time evolution of a classical statistical ensemble can be discussed independently of the Grassmann formulation (cf. sect. \ref{Evolutionlaw}). This also holds for the wave function. The role of the quantum wave function is best characterized by mapping the wave function $\psi(t)$ to a local-time probability distribution $\{p_\tau(t)\}$ on one side, and by a map from the all-time probability distributions $\{p_\omega\}$ to quantum wave functions on the other side. We start with the map $\psi(t)\to \{p_\tau(t)\}$. 

Consider first an arbitrary real quantum mechanical wave function $\psi(\tau;t)$ with $\tau$ a set of continuous or discrete variables as position and spin. For any given time $t$ this defines a classical statistical ensemble with probability distribution
\be\label{P2}
p_\tau(t)\equiv p(\tau;t)=\psi^2(\tau;t).
\ee
A Schr\"odinger equation for the time evolution of $\psi$ induces a time evolution law for the classical probabilities $p_\tau(t)$. It is of the unitary type \eqref{21A}, \eqref{21C}, \eqref{21D}. 

The restriction to real wave functions entails no loss of generality. Indeed, we can always choose a real representation of the quantum  wave function. A complex function $\psi(\tau';t)$ can be decomposed into real and imaginary parts $\psi(\tau',t)=\psi_R(\tau',t)+i\psi_I(\tau',t)$. Interpreting $\eta=(R,I)$ as a discrete index with two values, we label $\tau=(\tau',\eta)$ and consider the real function $\psi(\tau;t)$ defined by $\psi(\tau',R;t)=\psi_R(\tau',t),\psi(\tau',I;t)=\psi_I(\tau',t)$. The unitary evolution of $\psi(\tau';t)$ described by the Schr\"odinger equation turns then to a rotation for the evolution of the real wave function $\psi(\tau;t)=q_\tau(t)$ according to eq. \eqref{21C}. 

The opposite direction, namely the association of a quantum wave function $q_\tau(t)$ to a given time evolution of a classical probability distribution $p_\tau(t)$, requires some additional thought since $q_\tau(t)$ is fixed by $p_\tau(t)$ only up to a sign $s_\tau(t)$
\be\label{P2a}
q_\tau(t)=s_\tau(t)\sqrt{p_\tau(t)}.
\ee
The choice of $\{s_\tau(t)\}$ can be largely fixed if we require that the time evolution of $\{q_\tau(t)\}$ is given by a continuous and arbitrarily often differentiable rotation. Indeed, if $q_\tau(t)$ depends continuously on time for a given choice of $s_\tau(t)$, any other choice of $s_\tau(t)$ would lead to a discontinuous jump of $q_\tau$ whenever $q_\tau\neq 0$. For continuous $q_\tau(t)$ the sign $s_\tau(t)$ must be constant for all $t$ where $q_\tau(t)\neq 0$. However, $s_\tau(t)$ may change sign for the zeros of $q_\tau$. This typically happens if a continuous rotation implies a simple zero $q_\tau(\bar t)=0$, with 
$sign\big (q_\tau(\bar t+\epsilon)\big)=-sign\big(q_\tau(\bar t-\epsilon)\big)$. More complicated cases can be discussed in a similar spirit. These arguments fix $\{s_\tau(t)\}$ for every choice of the initial values $\{s_\tau (t=0)\}$. Any other choice would imply a discontinuity in the derivatives of the parameters of the $O({\cal N})$-rotations at some time $t$. Such discrete jumps are excluded if we require continuity and arbitrary differentiability of the time evolution for $q_\tau(t)$. 

The ``history of signs'' $s_\tau(t)$ stores information that is available for the sequence of local-time probability distributions $\{p_\tau(t)\}$, but not for an individual probability distribution at fixed $t$. These signs matter for the computation of expectation values of off-diagonal operators. The sequence of local-time probability distributions is computable from the all-time probability distribution $\{p_\omega\}$. We therefore have a map $\cP_l\to{\cal S}$, where ${\cal S}$ denotes the local-time subsystem with a given evolution law, as characterized by the possible sequences of local probability distributions $\{p_\tau(t)\}$ that obey the evolution law. (Here $\cP_l$ denotes the ensemble of all-time probability distributions for which the evolution law is valid.) For a unique construction of a wave function from classical probabilities we still need a map ${\cal S}\to{\cal Q}$, with ${\cal Q}$ the space of quantum states, parametrized here by the real wave functions $\{q_\tau (t)\}$. This association is still not unique, since the signs of the wave function $\{q_\tau(t=0)\}$ are not fixed uniquely. Only once the signs $s_\tau(t=0)$ are fixed the signs for all other times $s_\tau(t)$ are computable by the evolution law.

If the index $\tau$ contains a continuous  part, as a position variable $x$ or a momentum variable $p$, the choice of $s_\tau(t=0)$  may be further restricted by requesting continuity and differentiability of $q_\tau(t)$ in $x$ or $p$. One typically ends with a freedom of choice of a few signs for $q_\tau(t=0)$, and we will pick a fixed one as in eq. \eqref{88B} for fixed $s$. This choice may be associated with a choice of gauge which does not matter for the expectation values of observables. For such a suitable fixed choice we can then map a given time evolution of classical probabilities $\{p_\tau(t)\}$ to a real quantum mechanical wave function $\{q_\tau(t)\}$. This completes the map $\cP\to{\cal Q}$, which amounts to a construction of a quantum state in terms of classical probabilities. 

In turn, this wave function may find a complex representation $\psi_{\tau'}(t)$ if the time evolution is compatible with the complex structures, i.e. if the $O({\cal N})$-rotations belong to the unitary subgroup $U({\cal N}/2)$. Furthermore, the relevant observables of the problem should be compatible with the complex structure. In the next section we give an explicit example how such a complex structure arises from classical probabilities. 

Most familiar quantum systems are described by a complex wave function $\psi_{\tau'}$. We emphasize that the classical probabilities are not the absolute square $|\psi_{\tau'}|^2$. Specifying only the absolute squares discards part of the information contained in the classical probabilities $p_\tau$. Since $|\psi_{\tau'}|^2=\psi(\tau',R)^2+\psi(\tau',I)^2$ the specification of $|\psi_{\tau'}|^2$ can no longer resolve between $p(\tau',R)$ and $p(\tau',I)$ and therefore looses phase information. 

It is often stated that the characteristics of quantum physics is encoded in the ``physics of phases'' that are needed beyond the probabilities. In our approach, most of the phase information is again associated to probabilities. The components of the real wave function $q_\tau$ contain the information for probabilities $p_\tau$. These are twice as many real numbers as $|\psi_\tau'|^2$ and therefore account for most of the phase information. What is missing beyond the probabilities $p_\tau$ are only the signs $s_\tau$ and one could phrase that the quantum state involves the ``physics of signs'' beyond the probabilities. These signs are needed for the computation of expectation values of off-diagonal operators. We have seen that part of the sign information is indeed coming from information about the sequence of local-time probability distributions beyond a given time. Another part may be fixed by continuity, and the remainder corresponds to a gauge choice without impact on observations. 

\medskip\noindent
{\bf 2. \quad Fixed time statistical ensemble and quantum 

\quad~ state}

At this point it is useful to introduce the notion of the ``fixed time statistical ensemble''. It is specified by the real wave function $\{q_\tau(t)\}$ for a fixed time $t$. This contains somewhat more information than a fixed time classical statistical ensemble which would be characterized by the probability distribution $\{p_\tau(t)\}$, namely the signs $\{s_\tau(t)\}$. These signs permit the computation of expectations values of real off-diagonal observables, besides the diagonal ones which only involve $\{p_\tau(t)\}$. 

If we specify the states of the local-time subsystem by the evolution law and the wave function $\{q_\tau(t_0)\}$, the information contained in the fixed time statistical ensemble and the state of the local-time subsystem is the same. We only have made a change of point of view by focusing on a given time $t$, and taking the evolution law as a dynamical statement how the fixed time statistical ensemble changes with time. For both the state of the local-time subsystem and the fixed time statistical ensemble the relative sign between different $s_\tau$ that are not determined by the time evolution may be partly fixed by continuity and partly by gauge convention. 

If a suitable complex structure exists the fixed time statistical ensemble can be associated directly with a pure state of a quantum system, as specified by a complex wave function. We will occasionally use the naming ``classical wave function'' for the real roots of the probabilities up to a sign, and ``quantum wave function'' for the usual complex wave function. The two are related by the presence of a complex structure.

\section{Complex structure and quantum phases}
\label{Twocomponentspinor}

A simple example for a complex structure may be instructive. Let us consider a four state system $(B=2,{\cal N}=4)$ with time evolution of the classical probabilities according to
\ba\label{P5}
p_1(t)&=&p_{1,0},\nn\\
p_2(t)&=&\cos^2(\omega t)p_{20}+\sin^2(\omega t)p_{30}\nn\\
&&-2\cos(\omega t)\sin(\omega t)\sqrt{p_{20}p_{30}},\nn\\
p_3(t)&=&\cos^2(\omega t)p_{30}+\sin^2(\omega t)p_{20}\nn\\
&&+2\cos(\omega t)
\sin (\omega t)\sqrt{p_{20}p_{30}},\nn\\
p_4(t)&=&p_{40}.
\ea
Here $p_{\tau,0}$ are the initial probabilities at $t=0$, normalized by $p_{10}+p_{20}+p_{30}+p_{40}=1$. This evolution can be described by a differential equation for the real wave function 
$\{(q_\tau)\},p_\tau=q^2_\tau$,
\be\label{P3}
\partial_tq_1=\partial_tq_4=0~,~\partial_tq_2=-\omega q_3~,~\partial_tq_3=\omega q_2.
\ee

The quantum Hamiltonian of this four state system is purely imaginary
\be\label{P6}
H=\omega\left(\begin{array}{cccc}
0,&0,&0,&0\\
0,&0,&-i,&0\\
0,&i,&0,&0\\
0,&0,&0,&0\end{array}\right).
\ee
We observe the close analogy with sect. \ref{Two-state}, with $q_2$ and $q_3$ playing the role of $q_1$ and $q_0$ in eq. \eqref{117A}. However, we have now a four-state system with other possible observables. It admits a Grassmann representation different from sect. \ref{Two-state}. A suitable Grassmann basis is $\{g_\tau\}=(1,\psi_1,\psi_2,\psi_1\psi_2)$, with Hamiltonian
\be\label{P7}
{\cal H}=-i\omega\frac{\partial}{\partial \psi_1}\psi_2+i\omega\frac{\partial}{\partial \psi_2}\psi_1
=\omega\sum_{\alpha,\beta}\frac{\partial}{\partial\psi_\alpha}
(\tau_2)_{\alpha\beta}\psi_\beta.
\ee

The time evolution \eqref{P3} can be described in a complex basis
\be\label{P9}
\tilde q=
\left(\begin{array}{c}
q_1+iq_4\\q_2+iq_3
\end{array}\right)~,~
i\partial_t\tilde q=H\tilde q~,~H=\frac{\omega}{2}(\tau_3-1).
\ee
This corresponds to a spin in a constant magnetic field. The complex structure is specified by the definition of $\tilde q$. Complex conjugation corresponds to an involutive map where the sign of $q_3$ and $q_4$ is flipped. Multiplication by $i$ in the complex basis corresponds in the real basis to the map $q_1\to-q_4,q_4\to q_1,q_2\to-q_3,q_3\to q_2$. In the complex basis the Hamiltonian is now a real hermitean operator. 

The particle numbers ${\cal N}_\alpha=\frac{\partial}{\partial\psi\alpha}\psi_\alpha$ are represented as linear operators acting on the real wave function $(q_1,q_2,q_3,q_4)$ as
\be\label{P10}
N_1=diag(1,0,1,0)~,~N_2=diag(1,1,0,0).
\ee
In the complex basis \eqref{P9} no linear representation as complex $2\times2$ matrices exists. These observables are not compatible with the complex structure. This also holds for the occupation number of the second component
\be\label{158A}
m_2 =diag(0,1,0,0)=N_2(1-N_1).
\ee
This particular occupation number corresponds to $N$ in sect. \ref{Two-state}, replacing eq. \eqref{61A} by $\kl N(t)\kr=p_2(t)$. The total particle number $N_{tot}=N_1+N_2=diag(2,1,1,0)$ is conserved $([N_{tot},H]=0)$, but again it is not compatible with the complex structure. 

An example for an observable which is compatible with the complex structure is 
\be\label{159B}
D=diag (0,1,1,0)=(N_1-N_2)^2.
\ee
It is represented in the complex basis as $(1-\tau_3)/2$. This two-level observable takes the value one if $N_1$ and $N_2$ are different, and zero if they are the same. This ``difference observable'' commutes with the Hamiltonian $([D,H]=0)$ and is therefore conserved. In the complex basis the hermitean operators can be written as linear combinations of the Pauli matrices $\tau_k$ and the unit operator. In the real basis $(q_1,q_2,q_3,q_4)$ these operators are mapped to $\tau_k\to \tilde \tau_k$, 
\be\label{234A}
\tilde \tau_1=1\otimes\tau_1~,~
\tilde\tau_2=\tau_1\otimes\tau_3~,~\tilde\tau_3=\tau_3\otimes\tau_3,
\ee
while multiplication with $i$ in the complex basis appears as a matrix multiplication with $I$ in the real basis,
\be\label{234B}
I=-i\tau_2\otimes \tau_1.
\ee
One wonders if besides $\tilde \tau_3=1-2D$ also the operators $\tilde \tau_1$ and $\tilde \tau_2$ can find a physical interpretation. They correspond to off-diagonal observables.

Our construction of classical statistical ensembles that admit a unitary time evolution law can easily be extended to a higher number of bits $B$, using a purely imaginary hermitean Hamiltonian acting on a real wave function $\{q_\tau\},\tau=1\dots 2^B$. In the limit $B\to\infty$ the wave function can depend on continuous variables. For example, we may consider real wave functions in phase space, $q(x,p)$, with $x,p$ continuous position and momentum variables. A possible Hamiltonian reads 
\be\label{243}
H_W=-i\frac pm\partial_x+V\left(x+\frac i2\partial_p\right)-V\left(x-\frac i2\partial_p\right).
\ee
This Hamiltonian can be used for a description of a quantum particle in a potential $V$ in terms of a classical probability distribution in phase space \cite{CW1}. From the classical  wave function $q(x,p)$ we can define a Wigner function \cite{Wig,Moyal}
\ba\label{244}
\rho_w(x,p)=\int_{r,r',s,s'}
&q&\left(x+\frac r2,p+s\right)\\
&q&\left(x+\frac{r'}{2},p+s'\right)\cos(s'r-sr').\nn
\ea
It has all properties of the Wigner function in quantum mechanics (for details see ref. \cite{CW1}) and therefore accounts for all aspects of a quantum particle. In particular, the usual complex wave function can be constructed fro the Wigner function for the particular case of pure quantum states.

\section{Quantum field theory for fermions}
\label{Quantumfieldtheory}
Our description of fermions in sect. \ref{Fermions} can be used directly for an implementation of a quantum field theory for fermions in a Hamiltonian framework. (The associated Grassmann functional integrals are discussed in ref. \cite{CWQFT}.) For this purpose the index $\alpha$ labeling the different fermion species is written as a double index $\alpha=(\vec x,s)$, where $\vec x$ is a position coordinate and $s$ labels internal degrees of freedom for the fermions. (Alternatively we may choose $\alpha=(\vec p,s)$ for momentum space.) In distinction to a single quantum particle where $\tau=\vec x$, the space label $\vec x$ denotes here the Grassmann variables $\psi_\alpha$ and not the Grassmann element $g_\tau$. We work here with a complex Grassmann algebra. Similar to sect. \ref{Twocomponentspinor} it can be obtained by using a suitable complex structure in order to group two ``real Grassmann variables'' into a ``complex Grassmann variables''. (for details of suitable complex structures see ref. \cite{CWQFT}.)

We can define creation and annihilation operators for a fermion of species $s$ at $\vec x$
\be\label{QF1}
a_s(x)=\psi_s(x)~,~a^\dagger_s(x)=\frac{\partial}{\partial\psi_s(x)}.
\ee
They obey the anti-commutation relations
\ba\label{QF2}
\big \{a_s(x),a_{s'}(x')\big\}&=&\big\{a^\dagger_s(x)~,~a^\dagger_{s'}(x')\big\}=0,\nn\\
\big\{a^\dagger_s(x),a_{s'}(x')\big\}&=&\delta_{ss'}\delta(x-x').
\ea
For formal precision, we may imagine here a discrete space lattice, 
$\vec x=\vec n\epsilon,\vec n=(n_1,n_2,n_3),n_k\in {\mathbbm Z}$, such that $\delta(x-x')=\delta_{n_1,n'_1}\delta_{n_2,n'_2},\delta_{n_3,n'_3}$. The particle number for a species $s$ at position $\vec x$ reads
\be\label{QF3}
{\cal N}_s(x)=\frac{\partial}{\partial \psi_s(x)}\psi_s(x)=
a^\dagger_s(x)a_s(x).
\ee
The eigenvalues of ${\cal N}_s(x)$ are one or zero, with $\big({\cal N}_s(x)\big)^2={\cal N}_s(x)$. All particle numbers commute 
\be\label{QF4}
\big[{\cal N}_s(x),{\cal N}_{s'}(x')\big]=0.
\ee
For later purposes we define the operators
\be\label{QF5}
{\cal M}_{st}(x,y)=a^\dagger_s(x)a_t(y),
\ee
with ${\cal N}_s(x)={\cal M}_{ss}(x,x)$, and commutation relations
\ba\label{QF6}
&&\big[{\cal M}_{st}(x,y),{\cal M}_{s't'}(x',y')\big]=\\
&&\delta_{s't}\delta(x'-y){\cal M}_{st'}(x,y')-\delta_{st'}\delta(x-y')
{\cal M}_{s't}(x',y).\nn
\ea

One can change to a momentum basis by
\be\label{QF7}
\psi_s(x)=\int_p e^{ipx}\psi_s(p),
\ee
such that the annihilation and creation operators for a fermion with momentum $p$ become
\ba\label{QF8}
a_s(p)=\int_x e^{-ipx}a_s(x)~,~a^\dagger_s(p)=
\int_x e^{ipx}a^\dagger_s(x),
\ea
with commutation relation
\be\label{QF9}
\big\{a^\dagger_s(p),a_{s'}(p')\big\}=\delta_{ss'}\delta(p-p').
\ee
(Appropriate factors $2\pi$ are included in $\int_p=d^3 p/(2\pi)^3$ and 
$\delta(p-p')=(2\pi)^3\delta^3(p-p')$. Alternatively, both $\int_p$ and $\delta(p)$ can be formulated for discrete momenta on a torus.) The particle number for momentum $p$
\be\label{QF10}
\ns(p)=a^\dagger_s(p)a_s(p)=\int_{x,y}
e^{ip(x-y)}{\cal M}_{ss}(x,y)
\ee
obeys 
\be\label{QF11}
\big[\ns(p),{\cal N}_{s'}(p')\big]=0,
\ee
but it does not commute with $\ns(x)$,
\ba\label{QF12}
&&\hspace{-1.8cm}\big[\ns(p),{\cal N}_{s'}(x)\big]=\delta_{ss'}\int_y
[e^{-ip(x-y)}{\cal M}_{ss}(y,x)\nn\\
&&
-e^{ip(x-y)}{\cal M}_{ss}(x,y)\big].
\ea
The total particle number obeys
\be\label{QF13}
N=\sum_s\int_x{\cal N}(x)=\sum_s\int_p{\cal N}(p)
\ee
and commutes with both ${\cal N}(x)$ and ${\cal N}(p)$. 

We can define the position and momentum operators for a multi-fermion system with $\kl N\kr> 0$ as
\ba\label{QF14}
X_k&=&\kl N\kr^{-1}\sum_s\int_x x_k\ns(x),\nn\\
P_k&=&\kl N\kr^{-1}\sum_s\int_p p_k\ns(p)\nn\\
&=&\kl N\kr^{-1}\sum_s\int_x a^\dagger_s(x)
(-i\frac{\partial}{\partial x_k})a_s(x),
\ea
where $\partial a_s(x)/\partial x_k$ can be defined by the difference of operators at two neighboring points. They obey the commutation relation
\be\label{QF15}
[X_k,P_l]=i\frac{N}{\kl N\kr^2}\delta_{kl}.
\ee
For eigenstates of $N$ the commutator vanishes $\sim 1/N$ for $N\to\infty$. At this point we do not question about the basic origin of non-commuting operators and their association to observables. This issue concerns the all-time probability distribution which we do not address in this section. We simply note that all these operators act in a well defined way on any given Grassmann wave function $g(t)$. 

In particular, we may consider one-particle states. They correspond to eigenstates of total particle number with eigenvalue one,
\be\label{QF16}
Ng(t)=g(t).
\ee
The commutator $[X_k,P_l]=i\delta_{kl}$ takes now its standard form. It is straightforward to construct the eigenstates with $N=1$. For $S$ species fermions $(s=1\dots S)$ and $L$ lattice points $x,F=LS$, all contributions to $g$ must have precisely $F-1$ factors of $\psi_s(x)$. We may denote by $\tilde\psi_s(\vec x)$ the Grassmann elements to which $\psi_s(x)$ is conjugate,
\ba\label{QF17}
\int\cD\psi\psi_t(y)\tilde\psi_s(x)&=&(-1)^{F-1}\int\cD\psi\tilde \psi_s(x)\psi_t(y)\nn\\
&=&\delta_{st}\delta(x-y),
\ea
which obey
\ba\label{QF18}
\ns(y)\tilde\psi_t(x)&=&\delta_{st}\delta(x-y)\tilde \psi_t(x),\nn\\
N\tilde \psi_t(x)&=&\tilde\psi_t(x).
\ea
They can be obtained from the vacuum state
\be\label{QF19}
|0\kr=\psi_1(x_1)\psi_2(x_1)\dots\psi_s(x_1)\psi_1(x_2)\dots
\ee
(with some fixed ordering of the lattice points $x_i$) by
\be\label{QF20}
\tilde\psi_s(x)=\frac{\partial}{\partial\psi_s(x)}|0\kr.
\ee
(It is convenient to use here a different ``fermion convention'' for the choice of signs of the basis elements of the Grassmann algebra. We will take a plus sign for all ordered products of $\tilde \psi_s(\vec x)=\tilde\psi_\alpha$ for which the lower $\alpha$ are to the right.)

The most general eigenstate with $N=1$ then reads
\be\label{QF21}
g_1=\sum_s\int_x\varphi_s(x)\tilde\psi_s(x).
\ee
The coefficients of the states with fixed $\ns(x)=1$ can be associated with the components of a complex wave function $\varphi_s(x)$. It is normalized according to 
\be\label{QF22}
\sum_s\int_x\varphi^*_s(x)\varphi_s(x)=1,
\ee
such that the ``one-hole state'',
\be\label{176A}
h_1=\sum_s\int_x\varphi^*_s(x)\psi_s(x),\nn\\
\ee
is conjugate to $g_1$, i.e. $h_1=\tilde g_1$, or
\be\label{QF23}
\int\cD\psi h_1g_1=1.
\ee

For an arbitrary Grassmann operator ${\cal A}$ acting on $g_1$ we may define an associated operator $\hat A$ acting on $\varphi_s(x)$ by
\be\label{QF24}
{\cal A}g_1=\sum_{s,t}\int_{x,y}\hat A_{st}(x,y)\varphi_t(y)\tilde\psi_s(x).
\ee
This guarantees
\ba\label{QF25}
\kl A\kr&=&\int \cD\psi\tilde g_1{\cal A}g_1\nn\\
&=&\sum_{r,s,t}\int_{z,x,y}\varphi^*_r(z)\hat A_{st}(x,y)\varphi_t(y)\psi_r(z)\tilde\psi_s(x)\nn\\
&=&\sum_{s,t}\int_{x,y}\varphi^*_s(x)\hat A_{st}(x,y)\varphi_t(y),
\ea
which amounts to the standard quantum rule for expectation values in terms of associated operators. In particular, the position and momentum operators \eqref{QF14} are represented as
\be\label{QF26}
X_k=\delta(x-y)y_k~,~P_k=-i\delta(x-y)\frac{\partial}{\partial y_k}.
\ee
A simple proof uses the identities
\be\label{QF27}
\psi_s(x)\tilde\psi_t(y)=
\delta_{st}\delta(x-y)|0\kr~,~
\cD\psi|0\kr=1.
\ee
For one species $(S=1)$ we recover the quantum particle discussed at the end of the preceding section.

As a particular example for an evolution equation we may consider a system of electrons and positrons in an external electromagnetic field. The Hamiltonian reads $(S=4)$
\ba\label{QF28}
{\cal H}&=&\int_x\big\{m(\varphi^\dagger\varphi+\chi^\dagger\chi)
-eA_0(\varphi^\dagger\varphi-\chi^\dagger\chi)\nn\\
&&+i(\varphi^\dagger\tau_k\partial_k\chi^*+\chi^T\tau_k\partial_k\varphi)\nn\\
&&+eA_k(\varphi^\dagger\tau_k\chi^*+\chi^T\tau_k\varphi\big\}.
\ea
Here $\varphi$ and $\chi$ are two component Grassmann variables for quasi-electrons and quasi-positrons, respectively
\be\label{QF29}
\varphi=\left(\begin{array}{c}
\varphi_1(x)\\\varphi_2(x)
\end{array}\right)~,~\chi=
\left(\begin{array}{c}
\chi_1(x)\\\chi_2(x)\end{array}\right),
\ee
while $\varphi^*,\chi^*$ involve Grassmann-derivatives $\big(\varphi^\dagger=(\varphi^*)^T\big)$
\be\label{QF30}
\varphi^*=\left(\begin{array}{c}
\partial/\partial\varphi_1(x)\\ \partial/\partial\varphi_2(x)
\end{array}\right)~,~\chi^*=
\left(\begin{array}{c}
\partial/\partial\chi_1(x)\\ \partial/\partial\chi_2(x)\end{array}\right).
\ee
The space-derivative of a Grassmann-derivative is defined as
\be\label{QF31}
\partial_k\chi^*_1=\lim_{\epsilon\to 0}\frac{1}{\epsilon}
\left(\frac{\partial}{\partial\chi_1(x+\epsilon)}-\frac{\partial}{\partial\chi_1(x)}\right),
\ee
and $A_\mu=(A_0,A_k)$ denotes the electromagnetic potentials. 

The coinvariance of the corresponding evolution equation under Lorentz and gauge transformations is not easily visible in this formulation - the proof becomes straightforward for the associated Grassmann functional integral which is discussed in ref. \cite{CWQFT}. The last terms in eq. \eqref{QF8} annihilate or create simultaneously one quasi-electron and one quasi-positron. The total number of quasi-electrons plus quasi-positrons is not conserved. As is well known, there exists a modified total particle number which is conserved in the absence of electromagnetic fields. The Hamiltonian formulation of the time evolution can become rather cumbersome for strong fields where particle production plays a role. Many aspects are simpler in a functional integral formulation. Nevertheless, already at this stage one sees how a quantum field theory for Dirac fermions in an external electromagnetic field emerges from a classical probability distribution for discrete occupation numbers or Ising spins. 

\section{Conclusions}
\label{Conclusions}

Time can arise as a probabilistic structure for a suitable classical statistical ensemble. Since the concept of time is not introduced a priori, a probability distribution $\{p_\omega\}$ characterizing such an example cannot refer to any particular moment in time. It rather describes the complete possible probabilistic information about reality, covering the past, present and future, as well as possible situations for which time is only defined approximately and may loose its meaning under certain circumstances.

A time structure for $\{p_\omega\}$ requires an ordering structure for observables which permits to introduce the concept of local probabilities (local in time) by integrating out the past and future. The sequence of local-time probabilities $\{p_\tau(t)\}$ for different $t$ can be viewed as characterizing a state of a subsystem of the larger system described by the all-time probability distribution $\{p_\omega\}$, namely the local-time subsystem. While the all-time probability distribution $\{p_\omega\}$ contains, in principle, the information about all times, the local probability distribution $\{p_\tau(t)\}$ refers to a particular time. The expectation values of local observables $A(t)$ can be computed from $\{p_\tau(t)\}$ without requiring information concerning the past and future. If only the information about the local probability distribution is available or needed we deal with incomplete statistics \cite{3} for the local-time subsystem. 

Predictivity for the expectation values of observables at future times in terms of the expectation values at the present time requires the existence of an evolution law. Typically, such an evolution law is a differential equation for the time evolution of the local probabilities $p_\tau(t)$. For a given all-time probability distribution $\{p_\omega\}$ the evolution law (if it exists) can be computed. On the other hand, a particular evolution law is a particular structure of the all-time probability $\{p_\omega\}$ and restricts the possible $\{p_\omega\}$ which are compatible with this structure. 

We have presented explicit examples for all-time probability distributions $\{p_\omega\}$ which result in an oscillating evolution of the local-time probability distribution $\{p_\tau(t)\}$. We have also shown that an arbitrary time evolution of $\{p_\tau(t)\}$ can be implemented by a suitable $\{p_\omega\}$. Typically, many different $\{p_\omega\}$ can lead to the same time sequence of local probabilities $\{p_\tau(t)\}$. Such time sequences therefore define equivalence classes for all-time  probabilities $\{p_\omega\}$. The expectation values of the local observables as well as suitable correlations depend only on the equivalence class, such that the full information contained in a particular all-time probability distribution $\{p_\omega\}$ is not needed.

We concentrate on a suitable unitary evolution law which describes rotations of a real unit vector $q_\tau(t)$. The local probabilities are related to the ``wave function'' $q_\tau(t)$ by $p_\tau(t)=q^2_\tau(t)$, such that positivity, $p_\tau(t)\geq 0$, and normalization, $\sum_\tau p_\tau(t)=1$, are easily implemented. The sign $s_\tau(t)$ of $q_\tau(t)=s_\tau(t)\sqrt{p_\tau(t)}$ is largely fixed by the analytic properties of continuous rotations as well as possible analytic properties with respect to a continuous index $\tau$. Remaining ambiguities in the choice of $\{s_\tau(t)\}$ without physical relevance can be considered as a choice of gauge. The wave function $\{q_\tau(t)\}$ at a given time $t$, together with the evolution law contains essentially the same information as the sequence of local-time probabilities $\{p_\tau(t)\}$. In particular, the sequence $\{p_\tau(t)\}$ contains (partial) information about the distribution of sign functions $\{s_\tau(t)\}$ which is needed for the computation of expectation values of off-diagonal observables. 

It is more convenient, however, to consider the signs as a genuine part of the local-time information and take the wave function $\{q_\tau(t)\}$ rather than the probabilities $\{p_\tau(t)\}$ as the basic ingredient for the description of physics at a given time $t$. For any given classical statistical ensemble $\{p_\omega\}$ the wave function $\{q_\tau(t)\}$ is, in principle, computable (up to irrelevant gauge choices).  From $\{q_\tau(t)\}$ one can compute directly the expectation values of observables $A(t+\alpha)$ at times different from $t$. These observables are associated to off-diagonal operators, similar to the Heisenberg picture in quantum mechanics. 

The unitary evolution law constitutes a direct implementation of the quantum mechanical wave function in terms of the classical probability density $\{p_\omega\}$. The quantum wave function at a given time $t$ is directly related to the sequence of local probability distributions $\{p_\tau(t)\}$. The components of a real quantum wave function are simply given by $q_\tau(t)$. This is no limitation since a complex wave function $\psi_{\tau'}(t)$ can always be written as a real wave function with twice the number of components. The unitary evolution equation for the local probabilities becomes the Schr\"odinger equation for the wave function in quantum physics. All quantum rules for the computation of expectation values of observables apply. Quantum physics results from classical probabilities which admit a time structure with unitary evolution law. 

A further fundamental concept of quantum physics and quantum field theory arises naturally from our formulation of probabilistic time, namely fermions and Grassmann variables. We have labeled the states $\omega$ of the classical ensemble by the eigenvalues of two-level observables which correspond to yes/no questions. Correspondingly, the states $\tau$ for the local-time subsystem (with associated probabilities $p_\tau(t)$) can be labeled by the eigenvalues of local number operators $N_\alpha(t)$ which can take the values zero or one. Associating $N_\alpha(t)=1$ with a fermion of type $\alpha$ to be present at the given time $t$, and $N_\alpha(t)=0$ with no fermion present, the states $\tau$ are basis states for the quantum wave function $\{q_\tau(t)\}$ for a multi-fermion system in the occupation number basis. 

We have presented various explicit examples for a quantum time evolution arising from a unitary evolution law for a classical statistical ensemble. This covers two-state quantum mechanics as well as quantum particles in a potential. Furthermore, we also have discussed quantum field theories for fermions. The expectation values of all diagonal observables can be computed from the local-time probabilities $p_\tau(t)$ in the standard way for classical observables, i.e. by expressions linear in $p_\tau(t)$. The off-diagonal observables involve, in addition, the signs $s_\tau(t)$ and depend on $p_\tau(t)$ in a non-linear manner.

Open issues remain. We have shown by explicit construction that all-time probabilities $\{p_\omega\}$ exist for any given sequence of local probabilities $\{p_\tau(t)\}$, and that many different $\{p_\omega\}$ lead to the same sequence. However, we have not yet searched for interesting realizations of all-time probabilities that would shed light on the emergence of particular evolution laws. In this context, we also have not yet found a simple direct construction of the wave function $\{q_\tau(t)\}$ in terms of $\{p_\omega\}$, which would supplement the simple rule \eqref{12} for the local probabilities $p_\tau(t)$. 

Another question concerns the precise choice of the one-dimensional ordering structure that represents time. In principle, there could be many different such structures. The predictions for observations could depend on the choice of the time structure (cf. ref. \cite{6A} for discussions of a similar problem in a different context). The requirement of a unitary evolution law already restricts the choice of suitable ordering structures very severely. Furthermore, one often would like to describe the dynamics by excitations from a time invariant ground state for which time is associated to a Killing vector. On the other hand, for general relativity or other models with diffeomorphism symmetry many timelike hyperfaces can be defined, with different ``times'' related by coordinate transformations. It will be an interesting question to find out for which situations time is essentially unique up to coordinate transformations. One may suspect that this issue is closely related to the loss of a unique metric close to the Planck scale, which suggests the emergence of a unique time only after the big bang \cite{24}. 

The non-uniqueness of the all-time probability distributions $\{p_\omega\}$ which realize a given time sequence of probabilities $\{p_\tau(t)\}$ or wave function $\{q_\tau(t)\}$ raises another important question. One may ask which unequal time correlation functions can be defined such that they can be computed only in terms of the sequence of wave functions $\{q_\tau(t)\}$. Such correlation functions differ from the classical correlation function. They are rather robust with respect to the details of $\{p_\omega\}$ since they only depend on equivalence classes of time sequences $\{q_\tau(t)\}$. Furthermore, they only depend on equivalence classes of observables and are therefore robust with respect to the precise choice of observables as well. This robustness suggests that actual measurement correlations are based on conditional probabilities related to such correlations. Correlations of this type have been discussed in a euclidean setting in ref. \cite{3} and found to be related to non-commutative operator structures. The present paper advances these ideas since a unitary time evolution is implemented directly in terms of probabilities, and not only by analytic continuation as in \cite{3}. We discuss the appropriate correlation functions explicitly for simple systems. They are associated to quantum operators and the associated operator product. 

Finally, one may wonder about the origin of the omnipresence of non-commuting observables in quantum physics. We have shown previously \cite{QPSS} how non-commuting observables can arise from standard classical observables for a classical statistical ensemble that comprises a subsystem and its environment. The system observables correspond to equivalence classes which reflect identical properties of the system, but different properties of the environment. The lack of commutativity reflects that the joint classical probabilities are not properties of the equivalence classes, as characteristic for incomplete statistics. The non-commutativity is related to a new product structure involving the equivalence classes. It is different from the commuting classical product. This new product defines correlations different from the classical correlations. While the general principle for the emergence of non-commuting observables is  known and some examples for non-commuting operators have been discussed in this paper, the present work has made no systematic attempt to construct non-commuting observables from underlying classical observables. Only for some simple cases we have related the non-commuting quantum operators directly to classical observables in the all-time ensemble. 

Despite these open points we find it remarkable in which simple way the basic concepts of quantum physics, as the unitary time evolution of probability amplitudes, non-commuting operators or fermions, arise from the formulation of the concept of time in a general probabilistic setting of classical statistics.

\LARGE
\section*{APPENDIX A: EVOLUTION LAW WITH TRANSITION MATRIX}
\label{Evolutionlawwith}

\normalsize
In this appendix we demonstrate that an evolution law with a transition matrix \eqref{15} does, in general, not account for periodic probabilities. The main reason is the lack of time reversal symmetry. This does, however, not exclude particular evolution laws based on a class of transition matrices for which time reflection symmetry can be realized. 

The evolution law \eqref{15} entails a composition property
\ba\label{18}
p_\tau(t)&=&\sum_{\rho,\sigma}W_{\tau\rho}(t,t-\epsilon)
W_{\rho\sigma}(t-\epsilon,t-2\epsilon)
p_\sigma(t-2\epsilon)\nn\\
&=&\sum_\sigma W_{\tau\sigma}(t,t-2\epsilon)p_\sigma(t-2\epsilon).
\ea
This can be ``integrated'', and we obtain for $t_3\geq t_2\geq t_1 \geq t_0$ the matrix multiplication
\be\label{19}
W(t_3,t_1)=W(t_3,t_2)W(t_2,t_1),
\ee
with
\be\label{20}
W(t,t)=1.
\ee
We can then express $p_\tau(t)$ in terms of an ``initial'' probability distribution $p_\tau(t_0)$,
\be\label{21Aneu}
p_\tau(t)=\sum_\rho W_{\tau\rho}(t,t_0)p_\rho(t_0).
\ee
We may interpret the probabilities $p_\tau(t)$ as components of a vector in ${\mathbbm R}^{{\cal N}}$. The basis vectors of ${\mathbbm R}^{{\cal N}}$ are the ``classical states'' $\rho$ for which $p^{(\rho)}_\tau=\delta^\rho_\tau$. The transition matrices act then as operators in this space. 

In turn, each classical state is a sequence of of occupation numbers for $B$ bits, and we can characterize $W$ by its operation on the bits. Consider the example of two bits $(B=2, {\cal N}=4)$, where $\rho=\big[(1,1),(1,0),(0,1),(0,0)\big]$. We can compose $W$ as a sum of ``elementary bit operators'' $B^{(v)}$, 
\be\label{20a}
W=\sum_v\kappa_vB^{(v)}~,~v=1\dots({\cal N}^2-1).
\ee
A typical bit operator is the annihilation operator $B^{(1)}_-$ for bit $1$ which changes $n_1=1$ to $n_1=0$ and yields zero when applied to a classical state for which $n_1=0$. It maps $(0,0)\to 0~,~(0,1)\to 0~,~(1,0)\to (0,0)~,~(1,1)\to (0,1)$ and corresponds to 
\be\label{21}
B^{(1)}_-=\left(\begin{array}{cccc}
0,0,0,0\\0,0,0,0\\1,0,0,0\\0,1,0,0
\end{array}\right)~,~
B^{(1)}_+=\left(
\begin{array}{cccc}
0,0,1,0\\0,0,0,1\\0,0,0,0\\0,0,0,0
\end{array}\right).
\ee
The hermitean conjugate $(B^{(1)}_-)^\dagger=B^{(1)}_+$ is the creation operator for bit $1$. Other operators are the switch between two bits $(0,1)\leftrightarrow(1,0)$, the diagonal number operators $N_1=diag(1,1,0,0),N_2=diag(1,0,1,0)$ or their product $N_1N_2=diag(1,0,0,0)$. The coefficients $\kappa_v$ have to obey restrictions in order to obey the relations \eqref{16} \eqref{17}. An example is $W=B^{(1)}_-+B^{(1)}_+$, with $W^2=1$. 

The simple periodic oscillations of a local two-state probability $(B=1)$, as given by eq. \eqref{15L}, cannot be expressed by transition matrices obeying eqs. \eqref{16}, \eqref{17}. Indeed, the most general form of $W_{\tau\rho}$ for $B=1$ reads
\ba\label{24A}
W=\left(\begin{array}{ccc}
1-p&,&p\\p&,&1-p
\end{array}\right).
\ea
If we characterize $W_1,W_2$ by $p_1,p_2$, the product $W=W_1W_2=W_2W_1$ has the form \eqref{24A} with
\be\label{24B}
p=p_1(1-p_2)+p_2(1-p_1).
\ee
Thus $W_1$ does not admit an inverse except for $p_1=0,1$, since $p$ differs from zero for arbitrary $p_2$. An inverse of $W$ would be needed if we want to represent the time reflected evolution of the local probabilities - we note that eq. \eqref{15L} is invariant under $t\to -t$. 

If we want to realize a continuous evolution in the limit $\epsilon\to 0$ we should have for $W(t,t-\epsilon)$ an infinitesimal form $p=A\epsilon~,~A\neq 0$. The time evolution obeys then the differential equation
\ba\label{24C}
\partial_tp_1&=&\frac 1\epsilon\big(p_1(t)-p_1(t-\epsilon)\big)\nn\\
&=&\frac 1\epsilon\Big\{ (W_{11}-1)p_1(t-\epsilon)+W_{10}\big(1-p_1(t-\epsilon)\big)\Big\}\nn\\
&=&-2A\left(p_1-\frac12\right).
\ea
For $A>0$ this corresponds to an irreversible approach to the fixed point at $p_1=p_0=1/2$ rather than to an oscillation. Similar problems persists for a higher number of species $B$ and we conclude that an evolution law involving transition matrices is often not suitable for a description of oscillating local probabilities. There are exceptional cases where a linear law of evolution can describe periodic probabilities and we will mention a few of them in this paper.

\LARGE
\section*{APPENDIX B: EVOLUTION OPERATOR AND QUANTUM FORMALISM}
\label{Evolutionoperator}

\normalsize
{\bf 1. \quad Grassmann evolution operator}

In the Grassmann representation the close correspondence of the ``unitary evolution law'' given by eqs. \eqref{21A}, \eqref{21D} or \eqref{21C} to the time evolution in quantum mechanics can be made apparent. For this purpose we employ operators ${\cal U}(t,t_0)$ acting in the Grassmann algebra. They describe the time evolution of a Grassmann element $g(t)=\sum_\tau c_\tau(t)g_\tau$ according to
\be\label{42a}
g(t)={\cal U}(t,t_0)g(t_0)~,~\tilde g(t)=\tilde{{\cal U}}^T(t,t_0)\tilde g(t_0).
\ee
Here we define $\tilde{{\cal U}}^T(t,t_0)$ such that for arbitrary elements~$\tilde g,f$ of the Grassmann algebra one has
\be\label{36a}
\int {\cal D}\psi\tilde{{\cal U}}^T\tilde g f=\int {\cal D}\psi\tilde g
\tilde{{\cal U}}f.
\ee
The associated time evolution of $p_\tau(t)$ obtains for a given evolution operator ${\cal U}(t,t_0)$ as
\ba\label{43}
p_\tau(t)&=&\int{\cal D}\psi\tilde g(t) {\cal P}_\tau g (t)\nn\\
&=&\int{\cal D}\psi \tcu^T(t,t_0)\tilde g(t_0){\cal P}_\tau {\cal U}(t,t_0)g(t_0).
\ea

We may express the linear operators ${\cal U},\tcu^T$ as matrix multiplications
\ba\label{44}
{\cal U}(t,t_0)g(t_0)=\sum_{\tau,\rho} U_{\tau\rho}(t,t_0)c_\rho(t_0)g_\tau,\nn\\
\tcu^T(t,t_0)\tilde g(t_0)=\sum_{\tau,\rho} \tu^T_{\tau\rho}(t,t_0) c^*_\rho(t_0) \tilde g_\tau,
\ea
with
\be\label{44A}
\tu^T_{\tau\rho}=\tu_{\rho\tau},
\ee
such that
\ba\label{45}
p_\tau(t)&=&\int{\cal D}\psi \sum_{\lambda,\rho,\sigma} \tu^T_{\lambda\rho} c_\rho^* U_{\tau\sigma} 
c_\sigma  \tilde g_\lambda g_\tau\nn\\
&=&\sum_{\rho,\sigma}c^*_\rho(t_0)\tu_{\rho\tau}(t,t_0) U_{\tau\sigma}(t,t_0)c_\sigma(t_0).
\ea
From $\sum_\tau p_\tau =1$ we infer
\be\label{46}
\sum_{\rho,\sigma} (\tu U)_{\rho\sigma}c_\rho^*c_\sigma = 1,
\ee
which can hold for arbitrary $c_\tau$ obeying $\sum_\tau \vert c_\tau \vert^2=1$ only if $(\tu U)_{\rho\sigma}=\delta_{\rho\sigma}$. Thus the matrix $\tu$ is the inverse of $U$. Furthermore, the condition that $p_\tau(t)$ is real is obeyed by eq. \eqref{45} if $\tu^T= U^*$. Then $p_\tau(t) \geq 0$ follows automatically
\be\label{47}
p_\tau(t)=\vert\sum_\sigma U_{\tau\sigma}(t,t_0)c_\sigma(t_0) \vert^2.
\ee
We conclude that $U$ is a unitary matrix, $U U^\dagger=1$, $\tu=U^\dagger$. 
Already at this stage we recognize the unitary time evolution operator of quantum mechanics. The Grassmann operator ${\cal U}$ plays a role similar to the unitary evolution operator in quantum mechanics, 
\be\label{50}
\tcu (t,t_0) {\cal U}(t,t_0)= 1_{\cal G}.
\ee
For the case of real $c_\tau=q_\tau$ we have to require that $q_\tau=\sum_\rho U_{\tau\rho} q_\rho$ is real. This restricts the unitary transformations to the subgroup of real matrices $U$. Those are the rotations $SO({\cal N})$, or $U=R, R^T R=1$, and we recover eq.\eqref{21C} if we consider the $g_\tau$ as a fixed basis of the Grassmann algebra. 

We note that the Grassmann algebra admits an involution
\be\label{51}
g_\tau \leftrightarrow \tilde g_\tau~,~ c_\tau \leftrightarrow c^*_\tau  ~,~ g \leftrightarrow \tilde g.
\ee
This can be used in order to define the notion of a complex conjugation in the Grassmann algebra
\be\label{52}
(c_\tau g_\tau)^*=c_\tau^* \tilde g_\tau,
\ee
which is a nontrivial operation even for real $c_\tau = q_\tau$. In terms of this complex structure we can regard $\tcu$ as the hermitean conjugate of $\cal U$.

\medskip\noindent
{\bf 2. \quad Schr\"odinger equation}

The analogy to quantum mechanics is apparent if we interpret the Grassmann-valued wave function $g$ as a vector with components
\be\label{53}
\varphi_\tau(t)={\cal P}_\tau g(t)=c_\tau(t)g_\tau.
\ee
The complex conjugate wave function $g^*=\tilde g$ has components
\be\label{54}
\tilde\varphi_\tau(t)={\cal P}_\tau\tilde g(t)=c^*_\tau(t)\tilde g_\tau=
\varphi^*_\tau(t).
\ee
In terms of the components we can write
\be\label{55}
p_\tau(t)=\int\cD\psi\tilde\varphi_\tau(t)\varphi_\tau(t).
\ee
Expectation values of observables find an expression similar to quantum mechanics,
\be\label{56}
\kl A(t)\kr=\int \cD\psi\sum_\tau\tilde\varphi_\tau(t){\cal A}\varphi_\tau(t),
\ee
with ${\cal A}$ acting as a diagonal operator ${\cal A}\varphi_\tau=A_\tau\varphi_\tau$. 

In this picture the time evolution can be written as the usual unitary evolution of the wave function
\ba\label{57}
\varphi_\tau(t)&=&\sum_\rho U_{\tau\rho}(t,t_0)\varphi_\rho(t_0),\nn\\
\tilde\varphi_\tau(t)&=&\sum_\rho U^*_{\tau\rho}(t,t_0)\tilde\varphi_\rho(t_0)~,~U^\dagger U=1,
\ea
which yields explicitly
\ba\label{58}
p_\tau(t)&=&\int\cD\psi\tilde\varphi_\tau(t)\varphi_\tau(t)\nn\\
&=&\sum_{\rho,\sigma}U^*_{\tau\rho}(t,t_0)U_{\tau\sigma}(t,t_0)\int\cD\psi
\tilde\varphi_\rho(t_0)\varphi_\sigma(t_0)\nn\\
&=&\sum_{\rho,\sigma}U^*_{\tau\rho}(t,t_0)U_{\tau\sigma}(t,t_0)
\rho_{\sigma\rho}(t_0)\nn\\
&=&\big (U(t,t_0)\rho(t_0)U^\dagger(t,t_0)\big)_{\tau\tau}=\big(\rho(t)\big)_{\tau\tau}.
\ea
Here we have introduced the quantum mechanical ``density matrix''
\be\label{59}
\rho_{\tau\sigma}=\int \cD \psi\tilde\varphi_\tau\varphi_\sigma~,~
\rho=\rho^\dagger~,~ \text{tr}\rho=1,
\ee
with real diagonal elements $\rho_{\tau\tau}=p_\tau$ obeying $0\leq\rho_{\tau\tau}\leq 1$. 

The quantum  mechanical formalism can also be presented in the standard way by omitting the basis vectors $g_\tau$ and defining complex vectors $\varphi$ and $\varphi^*$ with components $c_\tau,c^*_\tau$, with associated density matrix $\rho_{\tau\sigma}=c_\tau c^*_\sigma$. The classical observables can be represented as diagonal operators $\hat A,(\hat A)_{\tau\sigma}=A_\tau\delta_{\tau\rho}$, and we can define the usual scalar product (with $|\varphi\kr=\varphi,\kl\varphi|=\varphi^*$, such that 
\be\label{60}
\kl A\kr=\kl \varphi|\hat A|\varphi\kr=\text{tr}(\rho\hat A).
\ee
Even though we describe a classical statistical ensemble, the unitary time evolution \eqref{21A} , \eqref{21C} can be fully described with the formalism of quantum mechanics. Of course, at this stage the observables all commute and are all represented by diagonal operators. Furthermore, the wave function $\varphi$ remains real if we use the real Grassmann algebra with $c_\tau=c^*_\tau=q_\tau$. 

The time evolution of the Grassmann wave function $g(t)$ obeys a differential evolution law
\be\label{94A}
i\partial_tg(t)={\cal H}(t)g(t)
\ee
with Hamilton operator ${\cal H}$ defined by eq. \eqref{113A}. In terms of the components of the wave function eq. \eqref{94A} becomes 
\ba\label{113C}
i\partial_t\varphi_\tau(t)&=&\sum_\rho H_{\tau\rho}(t)\varphi_\rho(t),\nn\\
i\partial_tc_\tau(t)&=&\sum_\rho H_{\tau\rho}(t)c_\rho(t).
\ea
This shows explicitly that the Schr\"odinger equation can be obtained from an appropriate evolution law for classical probabilities.

\end{document}